\documentclass[useAMS,usenatbib,aas_macros]{mn2e}

\usepackage{epsfig}
\usepackage{amssymb}
\usepackage{natbib}
\usepackage{aas_macros}

\DeclareGraphicsExtensions{.eps}

\newcommand{\bt}{$B/T$}
\newcommand{\PRe}{$P(R_e|M_{\rm star})$}
\newcommand{\vvir}{$V_{\rm vir}$}

\newcommand{\mbh}{$M_{\rm bh}$}
\newcommand{\mbhe}{M_{\rm bh}}

\newcommand{\mhe}{M_{\rm H}}
\newcommand{\rh}{$R_H$}
\newcommand{\re}{$R_e$}
\newcommand{\sis}{${\sigma}$}

\newcommand{\fgas}{$f_{\rm gas}$}

\newcommand{\mstar}{$M_{\rm star}$}

\def\msun{${\rm M_{\odot}}$}

\newcommand{\msune}{M_{\odot}}
\newcommand{\mstare}{M_{\rm star}}

\begin{document}

\def\sarc{$^{\prime\prime}\!\!.$}
\def\arcsec{$^{\prime\prime}$}
\def\arcmin{$^{\prime}$}
\def\degr{$^{\circ}$}
\def\seco{$^{\rm s}\!\!.$}
\def\ls{\lower 2pt \hbox{$\;\scriptscriptstyle \buildrel<\over\sim\;$}}
\def\gs{\lower 2pt \hbox{$\;\scriptscriptstyle \buildrel>\over\sim\;$}}

\title[Size Evolution]{Size Evolution of Spheroids in a Hierarchical Universe}

\author[F. Shankar et al.]
{Francesco Shankar$^{1}$\thanks{E-mail:$\;$francesco.shankar@obspm.fr},
Federico Marulli$^{2}$, Mariangela Bernardi$^{3}$, Simona Mei$^{4}$,
\newauthor
Alan Meert$^{3}$ and Vinu Vikram$^{3}$\\
$1$ Max-Planck-Instit\"{u}t f\"{u}r Astrophysik,
Karl-Schwarzschild-Str. 1, D-85748, Garching, Germany\\
$2$ Dipartimento di Astronomia, Universit\'{a} degli
Studi di Bologna, via Ranzani 1, I-40127 Bologna, Italy\\
$3$ Department of Physics and Astronomy, University of Pennsylvania,
209 South 33rd St, Philadelphia, PA 19104\\
$4$ GEPI, Observatoire de Paris, CNRS, Univ. Paris Diderot;
Place Jules Janssen, 92190 Meudon, France}

\date{}
\pagerange{\pageref{firstpage}--
\pageref{lastpage}} \pubyear{2012}
\maketitle
\label{firstpage}

\begin{abstract}
Unveiling the structural evolution of spheroids, and in particular
the origin of the tight size-stellar mass relation, has become one
of the hottest topics in cosmology in the last years and it is still
largely debated. To this purpose, we present and discuss basic
predictions of an updated version of the latest release of the
Munich semi-analytic hierarchical galaxy formation model that grows
bulges via mergers and disc instabilities.
We find that while spheroids below a characteristic mass
$M_s \sim 10^{11}\, \msune$ grow their sizes
via a mixture of disc instability and mergers,
galaxies above it mainly evolve via dry mergers.
Including gas dissipation in major mergers, efficiently
shrinks galaxies, especially those with final mass $M_s \lesssim 10^{11} \msune$
that are the most gas-rich, improving the match with different observables.
We find that the predicted scatter in sizes at fixed stellar mass
is still larger than the observed one by up to $\lesssim 40\%$.
Spheroids are, on average, more compact at higher
redshifts at fixed stellar mass, and at fixed redshift and stellar mass
larger galaxies tend to be more starforming. More specifically,
while for bulge-dominated galaxies the model envisages a nearly mass-independent decrease in sizes,
the predicted size evolution for intermediate-mass galaxies is more complex.
The $z=2$ progenitors of massive galaxies with $\mstare \sim (1-2) \times 10^{11}\, \msune$
and $B/T>0.7$ at $z=0$,
are found to be mostly disc-dominated galaxies with a median $B/T \sim 0.3$, with only
$\sim 20\%$ remaining bulge-dominated.
The model also predicts that central spheroids living in more massive haloes
tend to have larger sizes at fixed stellar mass.
Including host halo mass dependence in computing velocity dispersions,
allows the model to properly reproduce the correlations with stellar mass.
We also discuss the fundamental plane, the correlations with galaxy age,
the structural properties of pseudobulges, and the correlations with central black holes. 
\end{abstract}


\section{Introduction}
\label{sec|intro}

One of the most important and still debated problems in Cosmology is the
formation and evolution of galaxies. Today we see galaxies having a variety of morphologies,
ranging from less massive, pure stellar discs, to intermediate mass bulge plus disc galaxies,
to more massive, spheroidal systems. The origin of this transition is still for many respects unclear.
More specifically, while angular momentum conservation may explain many properties of discs \citep[e.g.,][]{Governato07},
the origin of bulges is still largely unsolved and debated.
Why do some galaxies show bulges while others don't?
Or, in other words, what is the origin of the gradual conversion from discs to spheroids?

It is clear that if we want to understand galaxy formation we need to observe
the high-redshift Universe. Deep observations in the last decades or so have
however unveiled a full complex zoology of high-redshift (proto)galaxies
that makes even more puzzling - but also more exciting - assessing the actual
routes chosen by nature to build the galaxy populations we observe today.
Along with starforming discs and dust-enshrouded galaxies \citep[e.g.,][]{Blain02,Magdis11},
deep optical and near-infrared surveys have in fact discovered the presence of numerous extremely compact and passively evolving galaxies up to $z\gtrsim 3$ \citep[e.g.,][]{Trujillo06,Trujillo07,Cimatti08,
Buitrago08b,Chapman08,Franx08,Saracco08,Tacconi08,Vanderwel08,Vandokkum08,Younger08,Bezanson09,Damjanov09,Williams09,Ryan10,Saracco10,VanDokkum10}.
The stellar masses range from $10^{10}-10^{12}\, \msune$, with a factor of $\sim 2$ systematic uncertainty, and with half-light radiuses within $0.4-5$ kpc, being $2-6$ times more compact than their local counterparts of similar stellar mass \citep[e.g.,][]{VanDokkum10,Saracco11}.
Although several observational limitations may affect high-redshift size measurements \citep[e.g.,][]{Mancini09}, extremely deep imaging \citep[e.g.,][]{Szomoru10} also through lensing \citep[e.g.,][]{Auger11,Newton11}, and the available measurements of very high velocity dispersions for a subset of these galaxies \citep[e.g.,][]{Cenarro09,Cappellari09,vandokkum09,Sande11}, are confirming their extreme compactness.

Moreover, it is now being established that early-type galaxies at higher redshifts are not all compact. At fixed stellar mass, similar fractions of large and compact galaxies of similar mass co-exist at the same epoch \citep[e.g.,][]{Mancini09,Valentinuzzi10a}, and with a variety of bulge-to-disc ratios \citep[e.g.,][]{vanderwel11}.
By studying the spectra of 62 early-type galaxies at high redshifts, \citet{Saracco11} found that compact galaxies tend to have most of their stars formed at $z>5$, while larger galaxies at fixed stellar mass are generally younger. Along similar lines and extending the analysis to other high-redshift galaxy populations, \citet{Mosleh11} concluded that the structure of galaxies is somewhat correlated to their activity, i.e.,
the sizes of galaxies at a given stellar mass is somewhat correlated to its star formation rate level, similarly to what is observed in the local Universe.
At lower redshifts it has been shown that the size-age relation at fixed stellar mass is similarly
shaped for lenticulars \citep{ShankarBernardi,vanderwel09}, i.e., older systems
are more compact, but becomes rather flat for bulge-dominated galaxies \citep{ShankarRe,Bernardi10,Trujillo11}. Thus whatever process formed massive ellipticals, it must have been fine-tuned to bring all young and old high-redshift massive spheroids on the same local size-mass relation.

Understanding the evolutionary link these compact and large high-redshift galaxies might have with the variety of starforming galaxies at similar redshifts and stellar mass, if any, and with the local early-type galaxy population remains an open debate \citep[e.g.,][]
{Cole00,Benson03,Granato04,Menci04,Baugh05,Bower06,Granato06,DeLucia06,Menci06,DeLucia07,Monaco07,Fan08,Somerville08SAM,Dekel09b,Khochfar09a,NeisteinWein,Bournaud11b,Gonzalez11}.

According to the standard cosmological paradigm of structure formation and evolution,
dark matter haloes have grown hierarchically, through the continuous merging of smaller
units into larger systems. In this scenario, galaxies form inside this hierarchically
growing system of haloes \citep[e.g.,][]{Cole00,DeLucia06}. However, the actual role
played by mergers (major, minor, wet, and dry), in the structural evolution of massive
spheroids is still uncertain \citep[e.g.,][and references therein]{HopkinsMergers}.
Some models of galaxy formation \citep[e.g.,][]{Eggen62,Merlin12} envisage that most
of the mass in local massive spheroids was formed \emph{and assembled} in a strong and rapid burst of star
formation at high redshifts, and the remnants evolved almost passively thereafter, without
being strongly affected by late merging events.

Galaxy formation models built on top of large N-body dark matter numerical simulations
or analytic merger trees, claim instead that although the stars of the most massive spheroids
are the oldest being formed at very high redshifts, they have assembled a large fraction of
their final stellar mass only at relatively late times via a sequence of minor and major merger
events \citep[e.g.,][]{Baugh05,DeLucia06,DeLucia07,Khochfar06a,Gonzalez11}.
It has long been known that binary mergers between discs can indeed produce
spheroidal galaxies and also explain many of their structural properties
\citep[e.g.,][]{Barnes92,Hernquist92,Robertson06}, though several issues
remain to be solved in this basic scenario \citep[e.g.,][]{Naab07}.
In high-redshift and gas-rich disc galaxy mergers, however,
gas dissipation inevitably forms compact spheroids \citep[e.g.,][]{Naab06,Robertson06,Hop08FP}.
Hierarchical models then naturally explain the evolution of the size (and mass)
of massive compact spheroids as a sequence of ``dry'' (gas-poor) and mainly
minor mergers that puff up the outskirts of the galaxy leaving the central regions
of the galaxy almost intact \citep[e.g.,][]{Naab07paper,Naab09,CiottiReview}.
Numerical simulations have however shed doubts on the
coherence with which mergers can bring galaxies along the tight structural relations
observed in the local Universe \citep[e.g.,][]{Ciotti01,Nipoti09}.

Another class of models explains size evolution of early-type galaxies
via a quasi-adiabatic expansion phase consequent to the blow-out
of substantial amounts of mass via quasar and/or stellar feedback \citep{Fan08,Damjanov09,Fan10}.
Initial numerical experiments to test the latter proposal as a viable explanation to size evolution
have been recently performed by \citet{RagoneGranato}.

Bulges could also be formed via in-situ processes that are broadly classified
as disc instabilities. In unstable self-gravitating discs,
the instability may drive the formation of a bar with
mass transferred from the disc into a central
bulge \citep[e.g.,][]{Cole00}. The degree of mass transferred
to the bulge varies from one model to the other.
Some models consider the instability quite a violent process
capable of transferring most of the disc into a bulge and also induce a starburst
\citep[][]{Bower06}. Besides bars, other types of instabilities could contribute to the
formation of bulges. Observations from the deep SINS survey of $z\sim 2$ galaxies
\citep{Genzel10,Forster11} have revealed the existence of gravitationally bound clumps
residing in gas-rich discs. Some hydrodynamic simulations have confirmed that
within the turbulent and gas-rich high redshift discs, large gas clumps
can indeed be formed and migrate via dynamical friction
to the centre thus progressively build a stellar bulge \citep[e.g.,][]{Dekel09a,Bournaud11a}.

Given the still debated physical mechanisms behind the evolution of spheroids, theoretical work on this hot topic is mandatory.
The aim of this work is to explore the full predictions of a state-of-the-art semi-analytic model (SAM) of
galaxy formation that evolves massive spheroids in a hierarchical fashion. More specifically,
we will present and discuss basic predictions on the size-mass relation and its evolution
with redshift of relatively massive spheroids. We will mainly focus on their size evolution
at fixed stellar mass and in different environments, the structural properties of their progenitors, but also
touch upon several other related issues.
Our objective is not to prove that the model discussed here is the correct one,
but rather to lay out the successes and failures of a detailed hierarchical model
against the wealth of data now becoming available from large and deep surveys. We will also
stress that several outcomes of the model considered here are shared by many other hierarchical SAMs,
making most of our conclusions of particular interest to the field of galaxy formation, but also discuss key differences.

The paper is organized as follows. In Section~\ref{sec|ModelsData} we introduce the hierarchical model considered for this work,
focussing on the features relevant to this work. In the same section we present a
large sample of well studied early-type galaxies taken as term of comparison for model outputs.
Section~\ref{sec|results} contains a discussion of key issues such as the size (and velocity dispersion)-stellar mass relation,
its scatter and evolution with redshift, and the role of environment.
We discuss in Section~\ref{sec|discu} the evolutionary features of spheroids, the analogies and
differences with other SAMs, and ways to further constrain galaxy formation models.
We then conclude in Section~\ref{sec|conclu}.
In the Appendices we will also briefly discuss a number of related topics,
such as the distinction between classical and ``pseudobulges'', the connection with black holes and with galaxy ages,
the fundamental plane and its evolution with time.

\begin{figure*}
    \includegraphics[width=15truecm]{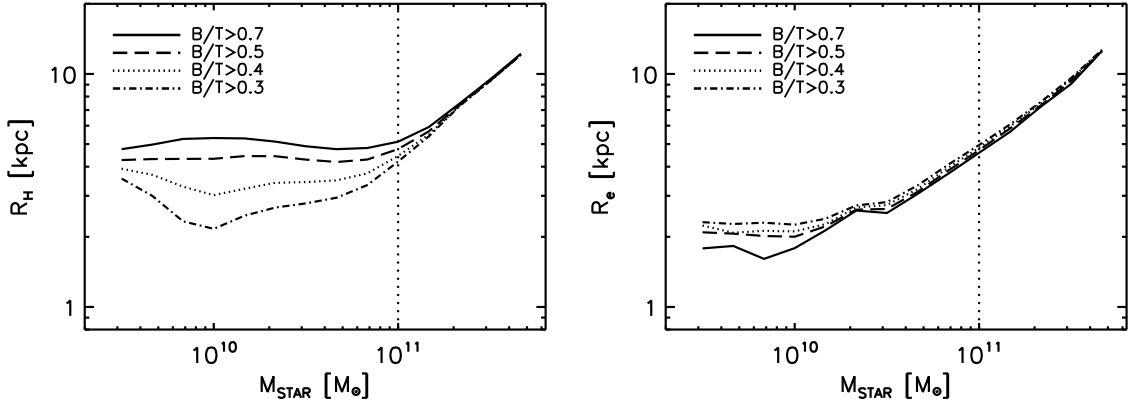}
    \caption{\emph{left panel}: Predicted median 3D half-mass radius versus
    stellar mass for different cuts in \bt, as labelled. \emph{Right panel}: median 2D projected half-light effective radius \re\
    as a function of stellar mass for a subsample of SDSS early-type galaxy sample.
    It is clear that the model is at variance with the data,
    predicting much flatter size-mass relations below a ``characteristic mass scale'' of
    $\mstare \sim 10^{11}\, \msune$, especially for higher \bt\ galaxies. This characteristic mass is completely absent in the data.}
    \label{fig|ReMsWithoutDissipation}
\end{figure*}

\section{Initial Settings: Model and Data}
\label{sec|ModelsData}

\subsection{The Reference Model}
\label{subsec|IntroRefMod}

All of the results presented in this work are the outcome of running the original
numerical source code by \citet{Guo11}, i.e., the latest rendition of the semi-analytic
model (SAM) developed at the Max Planck Institute
for Astrophysics. As detailed below, with respect to the original \citet{Guo11} model,
we have modified the computation of bulge radii exploring a variety
of possibilities, and added the calculation of the coupled velocity dispersions.
Note that running the code and producing a new galaxy catalog each time,
is different from studying the online catalogs as it allows a self-consistent
thorough study of the structural evolution of galaxies in the SAM.
We stress that the modifications applied to the original \citet{Guo11} model do not
affect any other galaxy property except for sizes. Thus the model (we checked)
maintains the same exact performance with respect to the
observables (e.g., the stellar mass function) as presented in \citet{Guo11}.
Before discussing bulge sizes in detail, we first provide below a brief overview of the model.

The Munich SAM aims at providing a comprehensive picture of the evolution of galaxies
and their central super-massive black holes within the hierarchical
structure and merging of dark matter haloes and subhaloes
within the concordance $\Lambda$CDM cosmology. To this purpose, it is
implemented on top of the large, high-resolution cosmological N-body
{\tt MILLENNIUM I} \citep{Springel05b} and {\tt MILLENNIUM II} \citep{Boylan09} simulations.
Given that in this paper we are mainly interested on the structural properties
of the most massive galaxies in the local Universe, all of the results presented
here have been obtained by running the code on the significantly larger
{\tt MILLENNIUM I} simulation.
The latter simulation follows the evolution of $N= 2160^3$ dark matter particles of mass
$8.6\times10^{8}\,h^{-1}{\rm M}_{\odot}$, within a comoving box of
size $500\, h^{-1}$Mpc on a side, from $z=127$ to the present, with cosmological parameters
$\Omega_{\rm m}=0.25$, $\Omega_{\rm b}=0.045$, $h=0.73$,
$\Omega_\Lambda=0.75$, $n=1$, and $\sigma_8=0.9$.

As comprehensively detailed in \citet{Guo11}, the model self-consistently evolves
the full population of galaxies within the hierarchy of dark matter haloes,
adopting a set of equations to describe the radiative
cooling of gas, the star formation, metal enrichment and supernovae
feedback, the growth and feedback of supermassive black holes, the
UV background reionization, and the effects of galaxy mergers.

Particularly relevant to the present paper is
the generation of galaxy morphology within the model.
Collapse of baryons within dark matter haloes and conservation of specific angular momentum naturally
leads to the formation of discs.
In the Guo et al. model gaseous and stellar discs are distinguished,
and each component is evolved in time in an inside-out fashion,
continuously evolving in mass and angular momentum due to the progressive
addition from cooling gas, minor mergers, and gas removal from star formation.
It is also assumed that both the gas and stellar discs
are thin, in centrifugal equilibrium, and to have exponential profiles
(we refer the reader to \citet{Guo11} for full details).

One of the primary channels
to instead form and evolve bulges in the model is via galaxy mergers.
In the case of a {\em minor} merger
($M_2/M_1<0.3$), the disc of the primary galaxy
survives, and the stars and the gas of the satellite are added to
any pre-existing bulge and to the disc of the primary galaxy, respectively.
Galaxy {\em major} mergers ($M_2/M_1\ge 0.3$)
instead disrupt any stellar disc present and produce a spheroidal remnant,
which contains all the old stars present in the progenitor galaxies and all the new stars formed
out of the burst triggered by the merger.

Bulges can be formed even via secular evolution in the model.
The type of disc instabilities considered in this model are secular
processes that transfer only the portion of the stellar mass necessary
to keep the disc marginally stable
(see details in Section~\label{subsec|RefModel} below, and in Guo et al. 2011).
This way of modelling disc instabilities
is different from, e.g., \citet{Bower06} that instead assume the entire mass
of the disc is transferred to the bulge during the instability, with any gas present
assumed to undergo a starburst. More in general, the present
model lacks at the moment any bulge formation via strong gas rich disc instabilities and/or clump
accretion under dynamical friction.

\citet{Guo11} have shown that their model is capable of reproducing
the size distribution of local discs reasonably well, and additional
comparisons can be found in, e.g., \citet{Fu10,Kauff12}. In this paper we will mainly focus on
the predicted structural properties of massive spheroids
and their evolution with redshift, and only briefly touch, where relevant, on the evolution of discs
that will be addressed elsewhere.
In particular, our primary interest in this work are bulge-dominated galaxies
with bulge-to-total ratio $B/T>0.7$,
a threshold chosen because, as discussed below, the contribution of pseudobulges
(Appendix~\ref{sec|sizemasspseudo}) in this regime is
negligible. This in turn allows us to properly
discern the actual role played by mergers in building
their structural properties we observe today. Where necessary, we will
also devote some attention to bulges grown via disc instabilities,
though we refer the reader to Appendix~\ref{sec|sizemasspseudo}
for some more specific discussion, and to the separate work by \citet{Shankar12}
for additional and complementary analysis of this issue.

\subsection{Computing bulge sizes in the Model}
\label{subsec|RefModel}

The bulge 3D half-mass radius \rh\ of a merger remnant is computed from energy conservation.
Following \citet{Cole00},
the model assumes that when two virialized galactic
systems merge, their \rh\
is given by
\begin{equation}
E_{fin}=E_{\rm int,1}+E_{\rm int,2}+E_{\rm orb}\, .
\label{eq|general}
\end{equation}

The left-hand side of Eq.~(\ref{eq|general}) is the
self-binding energy of the remnant, defined as
\begin{equation}
E_{\rm fin}=-\frac{G(M_1+M_2)^2}{R_H}\, ,
\label{eq|Efin}
\end{equation}
with \rh\ the half-mass radius of the remnant.

The terms $E_{\rm int, 1,2}$ are the self-binding energies
of the merging progenitors,
and are usually expressed as
\begin{equation}
E_{\rm int, i}=-\frac{GM_i^2}{R_i}\, ,
\label{eq|Eint}
\end{equation}
with $M_i$ the total stellar (including any stars formed during the merger) plus cold gas masses.

The orbital energy $E_{\rm orb}$ is usually expressed
in terms of the internal energy of the system
at the radius of minimal separation
\begin{equation}
E_{\rm orb}=-\frac{f_{\rm orb}}{c}\frac{GM_1M_2}{R_1+R_2}\, ,
\label{eq|Eorb}
\end{equation}
where we initially set $f_{\rm orb}=1$ and $c=0.5$ \citep[following][]{Cole00,Gonzalez09}.

Note that the conservation of energy in Eq.~(\ref{eq|general}) is not unique and
could possibly be expressed in other ways.
For example, the orbital energy of two galaxies at distance $d$ could also be
computed in the center of mass system of reference as
$E_{\rm orb}=\mu V^2/2-GM_1M_2/d$, in terms of the
reduced mass $\mu=M_1M_2/(M_1+M_2)$, and relative velocity $V=\parallel \vec{V}_1-\vec{V}_2 \parallel$.
However, while this expression might have the advantage of not being
directly dependent on $f_{\rm orb}$, it is still model-dependent. In fact,
satellite galaxies stripped away of their surrounding subhalo
are assigned a surviving merging timescale proportional to the actual Chandrasekhar dynamical friction timescale
via a fudge factor \citep[see][]{DeLucia10}. Thus relative velocities and distances would still need
to be modelled according to this timescale \citep[see also][]{Neistein11}.

Thus, given the inevitable inclusion of some parameters in the modelling of bulge sizes,
we decided for this work to stick with Eq.~(\ref{eq|general}) that requires
only one truly free extra parameter, $f_{\rm orb}/c$, and at the same time
allows a closer comparison with previous semi-analytic and numerical works in the Literature.

As anticipated in the previous
Section, another route to form bulges in the model is via secular evolution (Guo et al. 2011).
The adopted criterion for instability is expressed as $V_{\rm max}<\sqrt{GM_{\rm disc}/3R_{\rm disc}}$, with
$V_{\rm max}$ the maximum circular velocity
of the host subhalo, and $M_{\rm disc}$ and
$R_{\rm disc}$ the mass and exponential scalelength of the disc, respectively.
In the event of instability, a fraction $\delta \mstare$ of the disc stellar mass
is transferred to a central bulge to restore equilibrium, and the size $R_b$ of the newly formed bulge
is computed assuming an exponential profile. If a bulge is already present, the size of the bulge
is computed via a ``merger-type'' relation between the old and new bulge stellar mass as the one
in Eq.~(\ref{eq|general}), with $M_1$ and $R_1$ the mass and half-mass radius of
the pre-existing bulge, and $M_2$ and $R_2$ equal to $\delta \mstare$ and $R_b$, respectively,
and $f_{\rm orb}=2$ to take into account that the interaction in
concentric shells is stronger than in a merger (see Guo et al. 2011 for further details).

After each merger or disc-instability event, we also
compute the velocity dispersion associated to each galaxy following the
analytical fit given by \citet{Covington11}
\begin{equation}
\sigma^2=k \frac{G\mstare}{R_H}\left(1-\frac{M_H(<R_H)}{M_H(<R_H)+f(M_1+M_2)}\right)^{-1}\, ,
    \label{eq|CovingtonSigma}
\end{equation}
where we set $k=0.15$ and $f=0.1$. Here $M_H(<R_H)$ is the
fraction of the subhalo mass associated to the remnant within
the final half-mass radius.

\begin{figure*}
    \includegraphics[width=15truecm]{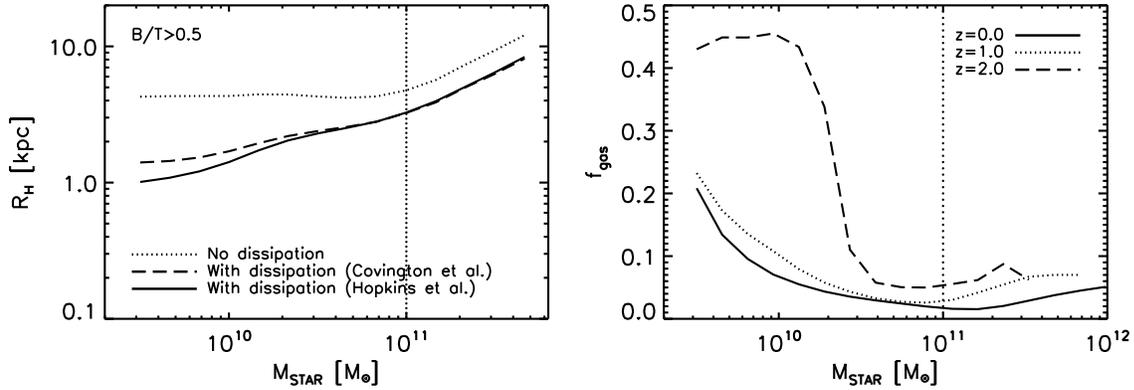}
    \caption{\emph{Left panel}: The \emph{dotted}, \emph{long-dashed}, and \emph{solid} lines
     show the predicted median size-mass relation for galaxies with $B/T>0.5$ for
     a model with no dissipation, with dissipation
    following \citet{Covington11}, and with dissipation following the prescriptions by \citet{Hop08FP}, respectively; to \emph{vertical dotted} line marks the transition
    above which (dry) mergers are believed to dominate galaxy assembly (see text). \emph{Right panel}:
    median cold gas fractions
    predicted in the model as a function of stellar mass at redshifts $z=0$, $z=1$, and $z=2$, as labelled.}
    \label{fig|ReMsWithDissipation}
\end{figure*}

\subsection{The SDSS sample}
\label{subsec|sample}

Throughout the paper we will compare model predictions with a sample of
$\sim 25,000$ galaxies from the Sloan Digital Sky Survey (SDSS). This is a
random subset of the well defined and complete sample defined in
Bernardi et al. (2010). This sample ranges between $3\times 10^9\, \msune$ to $\sim 10^{12}\, \msune$,
the mass range of interest here.
In addition to the photometric parameters (e.g., {\it cmodel} magnitudes and sizes)
presented in Bernardi et al. (2010), this subset of $\sim 25,000$ galaxies
also provides bulge-to-total $B/T$ light from a de Vaucouleurs \citep[e.g.,][]{deVac} plus exponential decomposition
(see Meert et al. and Vikram et al. in preparation), which we will extensively use in this
work to select the most bulge dominated galaxies to compare with the model.
The sample is characterized by stellar masses with a Chabrier Initial Mass Function \citep{Chabrier03},
consistent with the one used in the model.

\section{Results}
\label{sec|results}

\subsection{A first comparison}
\label{subsec|firstcomparison}

In Figure~\ref{fig|ReMsWithoutDissipation} we show a first comparison between model predictions and data from SDSS.
The left panel shows the median 3D half-mass radius \rh\ versus stellar mass for different cuts in \bt, as labelled.
From here onwards, unless otherwise stated, we will compute \rh\ as the mass-weighted average of the half-mass radiuses of the bulge
and the disc components. We have checked that, especially for the bulge-dominated systems of interest here,
this is equivalent (at the percent level) to computing a full mass profile assuming, e.g., an
Hernquist \citep{Hernquist90} plus an exponential profile for the bulge and the disc, respectively.
Moreover, simply neglecting the disc component in systems with $B/T \gtrsim 0.7$,
the subsample of galaxies this paper focuses on, yields very similar results.

The right panel shows the median effective radius \re\ as a function of stellar mass for the same cuts in
\bt\footnote{We note
that the behaviour of increasing half-mass radius with \bt\ at fixed mass is induced by the fact
that the model predicts larger bulge and disc sizes with increasing \bt. We verified
that just the opposite is true for disc-dominated galaxies with \bt$\lesssim 0.3-0.5$, in agreement
with observations.}.
The model predicts an increasing size with stellar mass, however,
it is evident that below a ``characteristic mass scale'' of $\mstare \sim 10^{11}\, \msune$
the predicted sizes flatten out, at variance with the data that continue to show a steep decline down to much lower masses
\footnote{The apparent flattening of the measured size-mass relation
at very low masses $\mstare \lesssim 10^{10}\, \msune$ is most
probably induced by contamination of later-type
galaxies, and thus not relevant for the present discussion
\citep[see details in][]{Bernardi11b}.}.

It is quite unlikely that such a strong discrepancy can be simply explained by some
mass/luminosity-dependent conversion factor between \rh\ and \re,
as we will also discuss in more detail later.
We checked that the predicted \rh-\mstar\ relation shows a very similar
flat behaviour at low masses already at $z\sim 2$.
The latter implies that the wrong shape of the size-mass relation
yielded by the model is a consequence of some wrong ``initial conditions'' and not necessarily
linked with any later galaxy assembly.

\subsection{Including Dissipation}
\label{subsec|includingdissipation}

If the merger is gas-rich, a significant
fraction of the initial energy $E_{\rm diss}$ of the system will be dissipated away,
inducing a more compact remnant. Several groups have studied this issue
in some detail using high-resolution hydrodynamic simulations and
semi-numerical models \citep[e.g.,][and references therein]{Naab06,Ciotti07,Hop08FP}.

Some of these groups have provided basic analytic formulations
that can be included in SAMs to study the impact of dissipation
on cosmological structure formation of galaxies.
One method is based on the conservation of energy, as proposed by \citet{Covington08}
\begin{equation}
E_{fin}=E_{\rm int,1}+E_{\rm int,2}+E_{\rm orb}+E_{\rm diss}\, ,
\label{eq|Covington}
\end{equation}
with the dissipation energy parameterized by \citet{Covington11} in terms of the final energy of the remnant as
\begin{equation}
E_{\rm diss}=2.75f_{\rm gas}E_{\rm fin}\, ,
\label{eq|Ediss}
\end{equation}
with $f_{\rm gas}$ the ratio between the total mass of cold gas
and the total cold plus stellar mass (inclusive of the mass formed during the burst) of the progenitors.

\citet{Hop08FP} parameterize the decrease in
size due to dissipation by the simple relation
\begin{equation}
R_H[final]=\frac{R_H[dissipationless]}{1+f_{\rm gas}/f_0}\, ,
\label{eq|Redissipation}
\end{equation}
with $f_0=0.25$, and $R_H[dissipationless]$ computed from Eq.~(\ref{eq|general}).

It is clear that Eqs.~(\ref{eq|Covington}) and (\ref{eq|Redissipation}) can significantly reduce the
sizes predicted in a dissipationless merger.
We assume that Eqs.~(\ref{eq|Covington})
and (\ref{eq|Redissipation}) only hold in \emph{major} mergers,
when the gas actually gets into the bulge (see Section~\ref{subsec|RefModel}).
The decrease is proportional to the gas fractions in the progenitors. As shown in the right panel
of Figure~\ref{fig|ReMsWithDissipation}, the model predicts increasing gas
fractions at lower stellar masses and, at fixed mass, increasing with redshift,
in broad agreement with observations \citep[e.g.,][and references therein]{Kanna04,Erb06,Catinella10,PeeplesShankar}.
The results shown here refer to galaxies with \bt$>0.5$
but they are general to galaxies with higher and somewhat lower \bt.
Low-mass galaxies can easily have most of their baryonic mass still in gaseous form at $z\gtrsim 2$. However,
at any epoch, galaxies above $M_c \gtrsim 3\times 10^{10}\, \msune$ tend to have progressively
lower gas fractions down to $\lesssim 10-20\%$. The mass $M_c$
is an interesting mass scale several times reported in the Literature
to be indicative of some basic
physical process in galaxy evolution (feedback from Active Galactic Nuclei?)
as several spectral and structural properties change when galaxies transition above it
\citep[e.g.,][]{Kauff04,Shankar06,Khochfar09a,Bernardi11a,Bernardi11b}.

This mass- and time-dependent behaviour of \fgas\ can then easily explain the
decrease in size shown in the left panel of Figure~\ref{fig|ReMsWithDissipation}.
The dotted, long-dashed, and solid lines show the predicted median size-mass
relation for galaxies with $B/T>0.5$ for a model with no dissipation, with dissipation
following \citet{Covington11}, and with dissipation following the prescriptions
by \citet{Hop08FP}, respectively.
Although a decrease in size is apparent at all masses,
dissipation is progressively more effective at lower masses, proportionally to the increase in gas fractions.
In both panels for reference the vertical dotted line at $\mstare = 10^{11}\, \msune$ marks the transition
above which dissipation does not play a significant role
in shaping the sizes of galaxies because mergers become progressively gas-poorer.
It is remarkable that dissipation tends to erase
the flattening below the characteristic mass producing a nearly single power-law
in agreement with the data.
In the following we will use Eq.~\ref{eq|Redissipation} as our reference model
with gas dissipation, though comparable results are obtained when switching to Eq.~\ref{eq|Covington}.

\begin{figure}
    \includegraphics[width=8.7truecm]{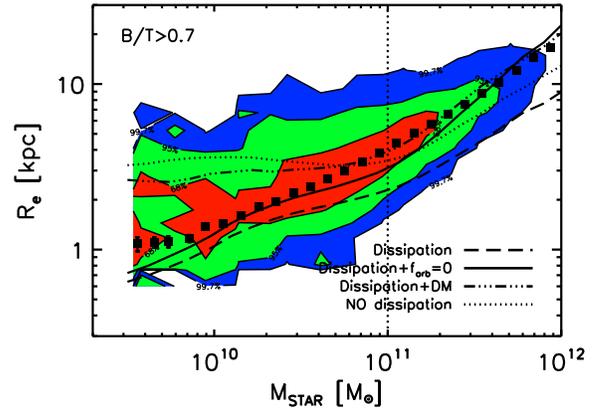}
    \caption{Effective radius \re\ as a function of stellar mass at $z=0$ from the SDSS
    sample of early-type galaxies with $B/T>0.7$. The contours mark the region of plane
    containing 68\%, 95\%, and 99.7\% fraction of the total sample. For completeness,
    the \emph{solid squares} represent the median size-mass relation for the early-type
    galaxy sample discussed by \citet{Bernardi11b}.
    The \emph{dotted}, \emph{long-dashed}, \emph{solid}, and \emph{triple dot-dashed} lines represent,
    respectively, the predicted size-mass relations without dissipation, with dissipation,
    with dissipation plus $f_{\rm orb}=0$, and with dissipation plus a fraction of dark matter in the merger (see text for details).}
    \label{fig|ReMsCompareWithData}
\end{figure}

\subsection{A closer comparison to the data}
\label{subsec|closercomparison}

We now attempt a closer comparison between model predictions and data measurements
by converting 3D half-mass radiuses \rh\ into 2D projected half-light radiuses \re.
Assuming that light traces mass we convert \rh\ to \re\ using the
tabulated factors from \citet{Prugniel97}. The latter computed
for a system of total mass $M$, the scaling factors $S(n)$,
dependent on the S\'{e}rsic index $n$ \citep{Sersic63}, connecting
the gravitational energy $W$ to their effective radius, i.e.,
$|W|=S(n)G M^2/R_e=GM^2/R_g$, with $G$ the gravitational constant and $R_g$
the gravitational radius.
Assuming the systems are virialized we can approximate $R_g\approx 2 R_H$, thus having
\begin{equation}
R_e\approx 2 S(n) R_H \, .
\label{eq|ReRhRelation}
\end{equation}
By setting $n=4$ in Eq.~(\ref{eq|ReRhRelation})
(i.e., $S(4)=0.34$ from Table 4 of \citealt{Prugniel97}),
we can convert the predicted 3D half-mass radiuses into
2D projected half-light radiuses\footnote{We have checked that our conclusions do
not significantly depend on the exact profile chosen for the bulges. For example,
we find broadly similar results, although somewhat steeper correlations at the highest
stellar masses, when assigning to each spheroid a
S\'{e}rsic index according
to their luminosity following, e.g., the empirical
relation by \citet{TerzicGraham}, and then converting from \rh\ to \re\
using the appropriate $S(n)$.}.

\begin{figure*}
    \includegraphics[width=15truecm]{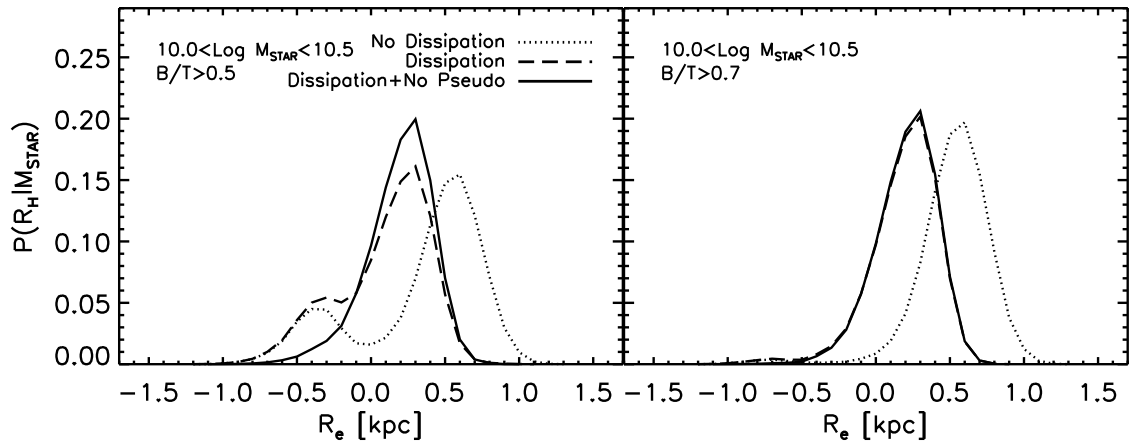}
    \caption{Relative size distribution of galaxies of a given mass and minimum \bt\
    as labelled.
    The \emph{dotted} and \emph{long-dashed}, and \emph{solid} lines are,
    respectively, for models with no dissipation, with dissipation including and excluding
    pseudobulges.
    More specifically, the \emph{solid} lines refer to models with dissipation
    but include only the subsample of galaxies grown mainly through mergers.
    Each distribution is
    normalized in a way that the sum of the galaxy fractions
    in each bin of size equals unity.}
    \label{fig|Scatter}
\end{figure*}

Figure~\ref{fig|ReMsCompareWithData} shows the
$z=0$ SDSS effective radius \re\ (assuming de Vaucouleurs plus exponential
profiles) as a function of stellar mass for galaxies with $B/T>0.7$, with the
contours marking the regions containing 68\%, 95\%, and 99.7\%
of the total sample (the results discussed below do not depend
on the exact choice of \bt\ threshold).  For completeness,
we report in the same Figure with solid squares
the median size-mass relation for the early-type
galaxy sample discussed by \citet{Bernardi11b}.

The lines in Figure~\ref{fig|ReMsCompareWithData} show the
predicted median size-mass relations for galaxies selected to have
the same minimum \bt\ threshold as in the data. The dotted line refers
to a model with no dissipation. As discussed above,
the latter model better lines up with the data for galaxies
above the characteristic mass, but below it the relation inevitably flattens out to larger sizes.
The long-dashed line
is the size-mass relation for a model that includes dissipation in major
mergers. As expected, the sizes get progressively smaller
towards lower mass spheroids that have formed
out of gas-richer progenitors.
However, spheroids of \emph{all} masses get shrunk.
Thus, including dissipation not only steepens the size-mass relation,
but it also lowers its overall normalization.

In order to improve the fit to the data with a model with dissipation we therefore need
to increase the normalization of the predicted sizes at fixed stellar mass.
Following Eq.~(\ref{eq|Efin}), we have that the size of the remnant is
$R_H\propto (M_1+M_2)^2/(E_1+E_2+E_{\rm orb}+E_{\rm diss})$,
so in order to increase the size at each merger event,
we need to either decrease the denominator,
and/or increase the numerator.
The solid line in Figure~\ref{fig|ReMsCompareWithData} is the
predicted size-mass relation assuming that
most of the merger events happen on parabolic orbits
with null orbital energy, i.e., $f_{\rm orb}=0$, a far from
uncommon condition in numerical simulations. Actually,
\citet{KhochfarBurkert} studied the orbital parameters of major mergers of
cold dark matter halos using a high-resolution cosmological simulation
finding that almost half of all encounters are nearly parabolic.
This simple variation to the basic model significantly
improves the match to the data.

We can however also increase the sizes by assuming that the total mass
actually taking part in the merger is the sum of the baryonic plus
a fraction of the dark matter host halo mass, i.e.,
\begin{equation}
M_i=M_{\rm star, i}+M_{\rm cold, i}+\alpha \times M_{\rm H}(<R_i)\,
\label{eq|Mi}
\end{equation}
where $R_i$ is the half-mass radius of the progenitor,
and $\alpha$ a constant parameterizing
the still uncertain effect of adiabatic contraction.
For each progenitor we take its mass at infall and compute the fraction
within the half-mass radius assuming a \citet{NFW} profile, and
assigning a concentration following the mean relation by \citet{Bullock01}.
The result is shown with a triple dot-dashed line in
Figure~\ref{fig|ReMsCompareWithData}, where we set $\alpha=1$.
The predicted sizes are larger, as expected,
though the inclusion of a constant fraction of dark matter also produces somewhat
larger sizes at low masses than actually observed (a similar behaviour
was discussed by \citealt{Gonzalez09}).

We conclude that dissipation inevitably shrinks galaxies
and thus some additional ingredient must be included
in the model to reestablish the normalization of the size-mass relation.
In the following we will use the model with $f_{\rm orb}=0$ and $\alpha=0$
as the reference one, unless otherwise stated.
We note, however, that including some
amount of dark matter participating in the merger
could still represent a viable model if we somehow tune $\alpha$ to
properly increase with halo/stellar mass\footnote{We stress here that a model
with $\alpha \ne 0$ in the merger cannot reproduce the tilt of the fundamental plane discussed in Section~\ref{subsubsec|FP},
that requires a halo mass-dependent velocity dispersion.}.

\subsection{On the scatter around the mean size-mass relation}
\label{subsec|scatter}

So far we discussed the median shape of the size-mass relation.
We now turn to the discussion of the dispersion in sizes
at fixed stellar mass. Figure~\ref{fig|Scatter} shows the relative
size distribution of galaxies in the mass range $10<\log \mstare <10.5$
and minimum \bt\ as labelled,
with each distribution \PRe\ normalized in a
way that the sum of the galaxy fractions in each size bin equals unity.

The dotted lines show the \PRe\ distributions competing to the no-dissipation model.
While galaxies with \bt$>0.7$ (right panel) have a Gaussian-like \PRe\ distribution,
galaxies with lower \bt$>0.5$ (left panel) tend to show a double-Gaussian distribution.
The second Gaussian is characterized by a peak and height a factor
of a few lower than the first Gaussian, though similar in amplitude.
The solid lines show the size distributions predicted by the model
with dissipation but removing those bulges that grew at least
50\% of their final size (and a large fraction of their mass)
via disc instability. The latter filtering is capable to
nearly erase the second lower Gaussian from the \PRe\ distribution,
proving that the latter is not produced by mergers but rather by disc instabilities.
We find that most of the remaining galaxies after the filtering have \bt$ \gtrsim 0.7$.

At masses higher than $\gtrsim 2\times 10^{11}\, \msune$ we always find
single-Gaussian distributions irrespective of the chosen \bt\ threshold.
We thus conclude that galaxies
of any mass with \bt$>0.7$ are predicted to possess bulges mainly
formed through mergers (the long-dashed and solid lines in the right panel
of Figure~\ref{fig|Scatter} coincide). On the other hand,
galaxies below the characteristics mass $M_s$
tend on average to have a significant population of pseudobulges with \bt$ < 0.7$.

\begin{figure}
    \includegraphics[width=8.7truecm]{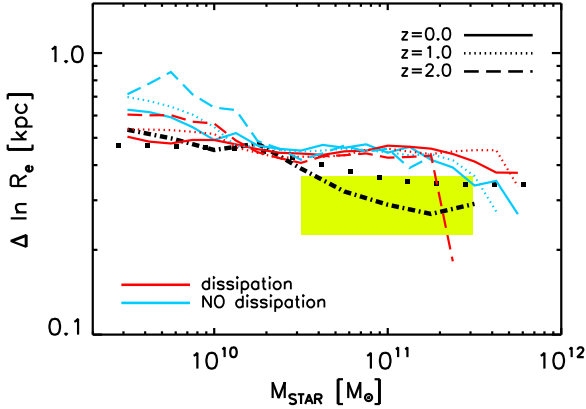}
    \caption{Predicted logarithmic $1-\sigma$ scatter in sizes at fixed stellar
    at different redshifts, as labelled.
    \emph{Red} and \emph{cyan} lines refer to model predictions
    with and without dissipation, respectively. For comparison,
    also shown the scatter measured in our SDSS subsample (\emph{thick}, \emph{dot-dashed} line),
    by Shen et al. (2003, \emph{filled squares}), and by Nair et al. (2011, \emph{yellow stripe}) for early-type galaxies.}
    \label{fig|ScatterRedshift}
\end{figure}

A full comparison between model predictions and observations is given
in Figure~\ref{fig|ScatterRedshift} where we plot the predicted
and observed scatter for galaxies with $B/T \gtrsim 0.7$
(so to minimize the contribution of pseudobulges in this regime.
We note, however, that even lower cuts
in \bt\ yield similar results, as long
as only classical bulges are considered).
We homogeneously compute the scatter in the data and in the
model by binning galaxies in stellar mass, build the
distribution in sizes and compute the 68\% percentile
as representative of its $1-\sigma$ dispersion.
Predictions for the models without and with
dissipation are shown with cyan and red lines, respectively,
at redshifts $z=0$, $z=1$, and $z=2$, as labelled.

The thick dot-dashed line is the scatter derived for
our SDSS galaxy subsample with the same cut in \bt.
We find it to be at the constant level of $\sim 0.5/\ln(10)\sim 0.22$ dex below $M_c$,
steeply decreasing to $\sim 0.1$ dex, the latter in very good agreement
with the results of \citet{Hyde09a} for high-mass early-type galaxies.
For completeness, the squared points represent the
scatter of early-type galaxies calibrated by \citet{Shen03}, while
the band marks the level of scatter measured, more recently,
by \citet{Nair11}. The latter computed
the scatter in sizes at fixed stellar mass
for early-type galaxies in rich environments.
The height of the band represents the uncertainty in
scatter they claim to find when measuring sizes in different ways.
Overall, despite the different techniques and selections
adopted by the different groups, all measurements agree quite well with each other.

The very first issue to note from Figure~\ref{fig|ScatterRedshift}
is that the predicted scatter is comparable to the observed one at low masses,
becoming progressively larger for galaxies with
$\mstare \gtrsim 3\times 10^{10}\, \msune$, the scale $M_c$ above which
galaxies become progressively gas-poorer and their structural evolution
becomes controlled by dry mergers.
The disagreement between model predictions and data
is mass dependent, and contained to be $\lesssim 40\%$.
This is in line with what claimed by several previous semi-analytic,
numerical, and observational studies that claimed the scatter
in size at fixed stellar mass to be in disagreement with the
observed one, especially for bulge dominated galaxies at high stellar masses
\citep[e.g.,][]{Gonzalez09,Nipoti09,ShankarPhire,Nair10,Nair11}.
This discrepancy represents a challenge for hierarchical models
that needs to be further understood.

When moving to significantly lower cuts in \bt\
the significant contamination from pseudobulges tends to further
increase the disagreement with observations.
We further note that the level of predicted scatter is only
marginally dependent on dissipation. In fact, galaxies with
stellar mass $\mstare \gtrsim 3\times 10^{10}\, \msune$ have
comparable levels of scatter in the two models, as expected
given that the role of (dry) mergers in the size evolution
becomes progressively more important at higher masses (see Section~\ref{subsec|evolution}).
The latter feature is in broad agreement
with some previous studies, though the role of
dissipation in determining the final scatter of sizes
was more emphasized \citep[e.g.,][]{KochfarSilk06Rez,Covington11}.
Also, the scatter does not strongly depend on redshift,
especially for more massive galaxies, again those
with $\mstare \gtrsim 3\times 10^{10}\, \msune$.

\begin{figure}
    \includegraphics[width=8.7truecm]{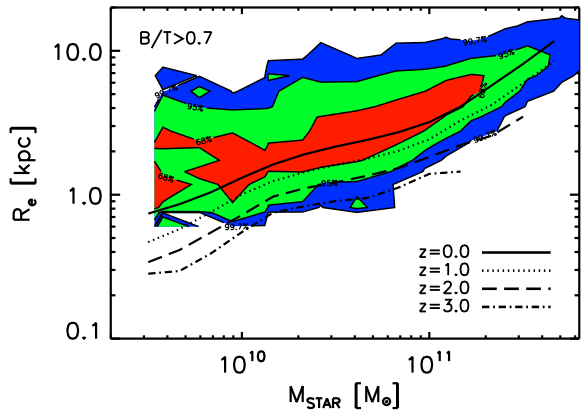}
    \caption{Predicted median size-mass relation for bulge-dominated galaxies with \bt$>0.7$,
    in terms of the projected radius \re, at different redshifts, as labelled,
    and compared to SDSS local data (coloured regions).
    There is a progressive decrease in sizes of up to a factor of $\sim 2$ at $z=3$,
    similar at all stellar masses.}
    \label{fig|ReMsRedshiftAll}
\end{figure}

\begin{figure}
    \includegraphics[width=7.5truecm]{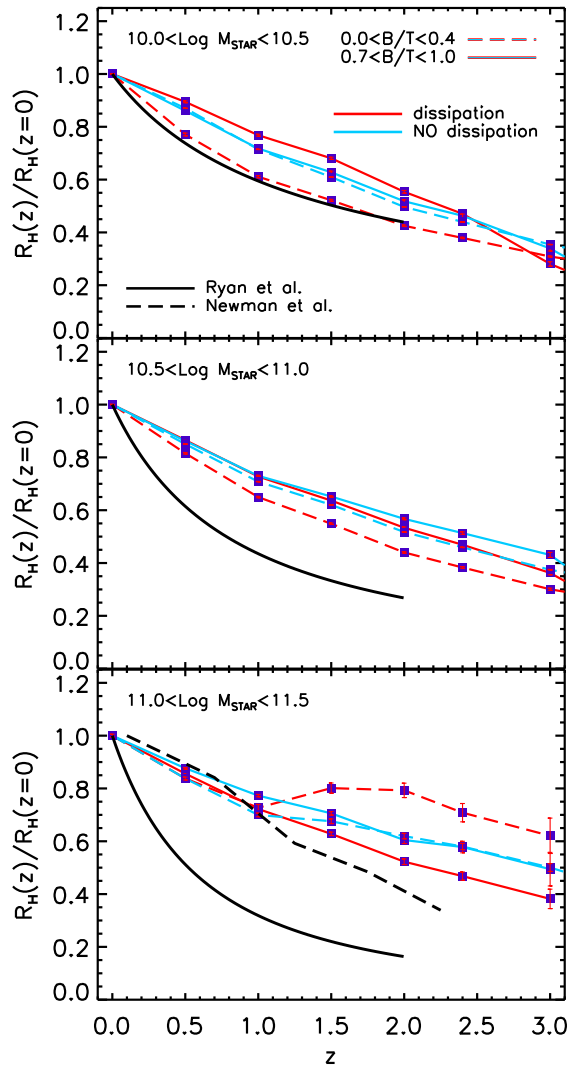}
    \caption{Predicted median redshift evolution (normalized to the median value at $z=0$)
    for the total half-mass radius of galaxies of different stellar mass as labelled and \bt\ as labelled.
    The \emph{upper}, \emph{middle} and \emph{lower} \emph{panels} refer to, respectively,
    the size evolution of low, intermediate,
    and high stellar mass galaxies divided into two intervals of \bt, as labelled.
    For bulge-dominated galaxies, the redshift evolution does not depend much on
    stellar mass or the degree of gas dissipation. Massive galaxies tend to evolve
    slower than what the data suggest (\emph{black}, \emph{solid} and \emph{long-dashed} lines).}
    \label{fig|ReMsRedshiftAll}
\end{figure}

\subsection{Size-evolution with redshift}
\label{subsec|evolredshift}

As discussed in Section~\ref{sec|intro} massive and passive spheroids
at high redshifts appear more compact with respect to their
local counterparts in SDSS. This section is dedicated to understand
the degree of size evolution at fixed stellar mass predicted by the
model and discuss it in light of the available data.

Figure~\ref{fig|ReMsRedshiftAll} shows the predicted
size-mass relation from our reference model,
in terms of the projected radius \re, at different redshifts,
as labelled, and compared to SDSS local values.
Here at all redshifts we select only bulge-dominated
galaxies with $B/T>0.7$. We find a progressive
decrease of sizes at all masses.
Galaxies with stellar masses above $\mstare \gtrsim M_c = 3\times 10^{10}\, \msune$
seem to experience a similar degree of evolution, i.e., a
progressive average decrease in size up to $\lesssim 3$ at $z \lesssim 3$.
This type of nearly mass-independent evolution implies that
the slope of the size-mass relation is predicted to be almost
constant at all redshifts, at least for bulge-dominated systems
with $B/T>0.7$ with $\mstare \gtrsim M_c$.

Figure~\ref{fig|ReMsRedshiftAll} presents a more specific
study of the redshift evolution in the median half-mass radius \rh,
normalized to the median local value, for two different intervals
of bulge-to-total ratios, i.e., \bt$<0.4$ and \bt$>0.7$, limits chosen
to select statistically significant
disc- and bulge-dominated galaxy samples, respectively. Each point
in the Figure represents normalized median values with their associated error bars.
The upper panel shows the evolution competing to lower mass galaxies
with $10^{10} < \mstare/\msune < 3\times 10^{10}$, the middle panel
for galaxies with $3\times 10^{10} < \mstare/\msune < 10^{11}$,
while the lower panel refers to higher mass galaxies
with $10^{11} < \mstare/\msune < 3\times 10^{11}$.
Globally, for all galaxies, even for the bulge-dominated ones (solid lines),
we do not find strong dependence of size evolution on gas
dissipation (red and cyan lines refer to model outputs with and without dissipation, respectively).
We have also checked that the degree of evolution in bulge sizes does
not significantly depend on the amount of orbital energy included in the model
(Eq.\ref{eq|general}). This is not unexpected given that the degree of bulge size evolution in
hierarchical models is mainly governed by the number and type
of mergers (Section~\ref{subsec|evolution}). In other words, dissipation mainly acts in deciding
how compact spheroids appear after the initial gas-rich major merger event,
leaving the degree of evolution, controlled by other processes,
not significantly perturbed.

From Figure~\ref{fig|ReMsRedshiftAll}
it is evident that the model, as anticipated in Section~\ref{subsec|IntroRefMod},
can manage to reproduce the moderate redshift
evolution in the half-mass stellar radius of disk-dominated galaxies (long-dashed lines), which
are observed to decrease in size by a factor of $\sim 2$ up to $z=2$ \citep[see, e.g.,][and references therein]{Somerville08}.
On average, however, we do not find the empirical trend for which
bulge-dominated galaxies tend to grow faster than disc-dominated ones
\citep[e.g.,][]{Somerville08,Ryan10,Huertas12}. If anything, we start seeing
a flatter size evolution in disc-dominated galaxies only for
galaxies above $\gtrsim 10^{11}\msune$ and at $z \gtrsim 1$.


\begin{figure*}
    \includegraphics[width=15truecm]{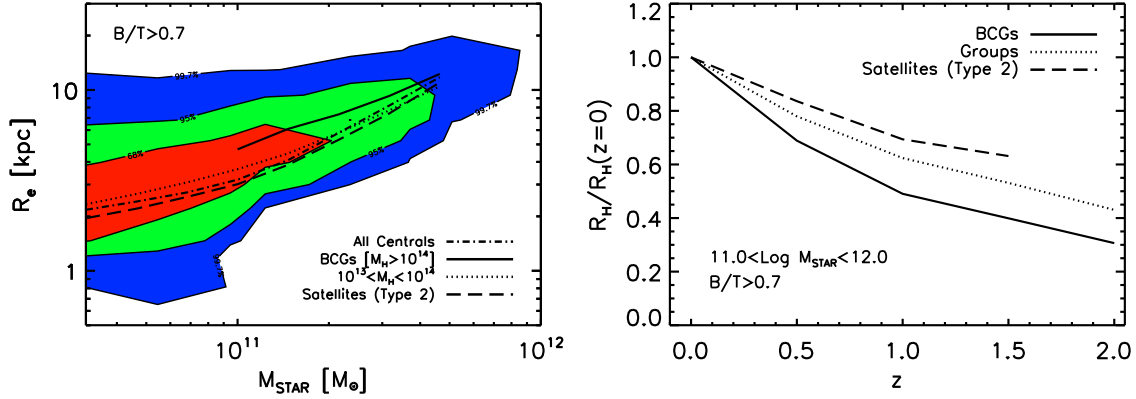}
    \caption{\emph{Left panel}: Predicted median size-mass relation
    for all central (\emph{dot-dashed} line) and Type 2 satellite
    galaxies  (\emph{long-dashed} line; with no
    restriction in host halo mass),
    compared to SDSS data. A systematic difference of
    $\sim 30\%$ with centrals being larger than satellites,
    is apparent at all masses. The \emph{dotted} and \emph{solid}
    lines refer to the size-mass relation of centrals in haloes with
    mass $10^{13}\,  < \mhe/\msune< 10^{14}$ (Groups) and $\mhe > 10^{14}\, \msune$ (BCGs),
    respectively.
    Central galaxies residing in more massive haloes tend be larger.
    \emph{Right panel}: median fractional size evolution for BCGs, centrals in groups,
    and Type 2 satellites. BCGs tend to have a much faster evolution
    than all other galaxies of similar mass.}
    \label{fig|ReMsType}
\end{figure*}

More generally, the model does not predict
a very strong size evolution for bulge-dominated
galaxies at fixed stellar mass, which might not be in full agreement with the data.
From the observational point of view in fact, despite the complexities
in the different selection processes \citep[e.g.,][]{VanDokkum10,vanderwel11,Huertas12,Newman12},
passive early-type galaxies tend to show a stronger size evolution at high stellar masses,
decreasing by up to a factor of $\lesssim 5$ \citep[e.g,][]{VanDokkum10,Huertas12}.
We report in each panel of Figure~\ref{fig|ReMsRedshiftAll} the analytic fit
of normalized size evolution at fixed stellar mass empirically derived by \citet[][black solid lines]{Ryan10},
based on direct deep $H$ imaging and data from the Literature.
The model predictions are in good agreement with the empirical fit
for masses below $\sim 3\times 10^{10}\, \msune$,
where the inferred size evolution is weaker,
but tend to progressively depart from the data at higher masses
(we stress that this comparison
is still at a qualitative level as the fits to observations have not been derived
for a homogeneous sample of galaxies selected to have the same \bt\ as in the model).

Similar results on the possible inefficiency of mergers in puffing up
massive galaxies have now been claimed also by several independent works
\citet{Cimatti12,Huertas12,Nipoti12}. \citet{Huertas12} more recently claimed
evidence for a weaker size evolution at intermediate masses $3\times 10^{10}<\mstare/\msune<10^{11}$,
in better agreement with model predictions,
but tend to confirm the strong drop in sizes for the highest stellar mass bins.
For completeness, we also report in the bottom panel of Figure~\ref{fig|ReMsRedshiftAll}
the fit recently inferred by \citet{Newman12} (long-dashed line) that better lines up with the
model predictions, at least at $z \lesssim 1$. We caution, however,
that due to their broader selections,
their sample may not be restricted to only early-type, bulge-dominated galaxies
\citep[see discussion in][and references therein]{Huertas12}.

In our study of size evolution at fixed stellar mass
we also tried to separate spheroidal galaxies that have mainly grown their
bulges via mergers from those that mainly grew their bulges via disc instabilities.
We found tentative evidence for pseudobulges to evolve slower in sizes with respect
to classical bulges of similar stellar mass, but the statistics in some bins is poor
and the systematic difference is confined at the $\lesssim 20\%$ level. SAMs that adopt
stronger disc instabilities could provide different conclusions in this respect.

\subsection{Role of Environment}
\label{subsec|ReMsType}

Not all early-type galaxies may follow the same size-mass relation. Environment,
or simply the special location of galaxies at their formation epoch,
might induce different evolution
at later times in galaxies of similar mass.
For example, if mergers dominates the structural growth
of galaxies, at least at lower redshifts, then galaxies
in denser environments where mergers are more efficient
might appear larger at fixed stellar mass.
From the observational point of view this is still debated.
At high redshifts, while some groups find clear evidence for larger galaxies
in denser environments, at fixed stellar mass \citep[e.g.,][]{Cooper12,Papovich11},
other don't or claim some stellar mass dependence \citep{Huertas12,Raichoor12}.
In the local Universe, when selecting galaxies of a given stellar mass in the field
and in overdense regions such as Clusters,
two main issues have emerged recently. Galaxies in Clusters less
massive than $\sim 4 \times 10^{11}$ M$_{\odot}$ tend to appear
smaller at fixed stellar mass than their local counterparts in the
field \citep{Valentinuzzi10a}, and non-central cluster galaxies
might have had a slower or even negligible evolution down to
$z=0$ \citep{Valentinuzzi10b,Saracco10}.
The observational evidence though is still sparse or not secure.
\citet{Weinmann09} investigated size distributions, among
other properties in the SDSS Data Release 4 group catalogue of \citet{Yang07},
finding no clear difference in the sizes of early-type centrals and satellites.
Even the degree of evolution for the brightest cluster galaxies (BCGs),
is not yet well understood.
\citet{Bernardi09} found that BCGs have evolved by 50\% in size in the last few Gyrs,
while \citet{Stott11} claim a milder evolution of $\gtrsim 30\%$ since $z=1$.
\citet{Ascaso11} also claim evidence for significant size evolution though
not in the light profile (the measured S\'{e}rsic index is nearly invariant with time).

Given the non-trivial impact that environment might induce on galaxy
size evolution, it is thus mandatory to study what the model predictions
are with respect to this important issue.
We recall that the \citet{Guo11} model follows in great
detail the fate of the gas and stellar components of galaxies
becoming satellites in larger dark matter haloes. Satellites
suffer tidal and ram pressure stripping that can remove
a large part of their gas reservoir. In particular,
the model further assumes that when the host subhalo
is completely disrupted during its journey within the largest halo,
the galaxy's stellar component starts also being disrupted
(see \citealt{Guo11} for details).
The stars which are stripped away can then later become part
of the central galaxy of the parent dark matter halo hosting
the satellite galaxy. Given all of these physical prescriptions,
it is natural to expect some structural differences
between central and satellite galaxies in the code.

The left panel of Figure~\ref{fig|ReMsType} shows the predicted
size-mass relation for central and satellite galaxies
(dot-dashed and long-dashed lines, respectively). Here satellites
are only the ones defined to be ``Type 2'' in the code, i.e.,
the ones that have completely lost their
associated subhalo due to disruption.
We find a relatively small, although systematic, difference with satellites
being smaller by $\lesssim 30\%$ with respect to centrals of the same stellar mass,
somewhat in between the findings of \citet{Weinmann09} and \citet{Valentinuzzi10a}.
The dotted and solid lines refer to the size-mass relation of
centrals in haloes with mass $10^{13}\,  < \mhe/\msune< 10^{14}$ and
$\mhe > 10^{14}\, \msune$, respectively. Central galaxies
residing in more massive haloes tend to be larger mainly
because they have undergone a larger number of mergers over cosmic time.

We have also analyzed size evolution for galaxies living
in different environments.
The right panel of Figure~\ref{fig|ReMsType} shows the
median fractional size evolution for galaxies at
fixed stellar mass and environment.
We consider galaxies having
stellar masses above $\gtrsim 10^{11}\, \msune$ and $B/T>0.7$.
We consider BCGs, identified as
centrals in haloes with $\mhe > 10^{14}\, \msune$ (solid line),
centrals in galaxy groups with $10^{13} < \mhe/\msune < 10^{14}$ (dotted line),
and Type 2 satellites galaxies with no restriction in
host virial mass (long-dashed line).
The model predicts that, at fixed stellar mass, galaxies residing in progressively
more massive haloes have a proportionally stronger size evolution
evolution, mainly induced by the larger number of mergers.
In particular, BCGs are predicted to increase in size by $\gtrsim 50\%$
at $z<1$, a degree of evolution in between
the one calibrated by \citet{Stott11} and \citet{Bernardi09}.

\begin{figure*}
    \includegraphics[width=15truecm]{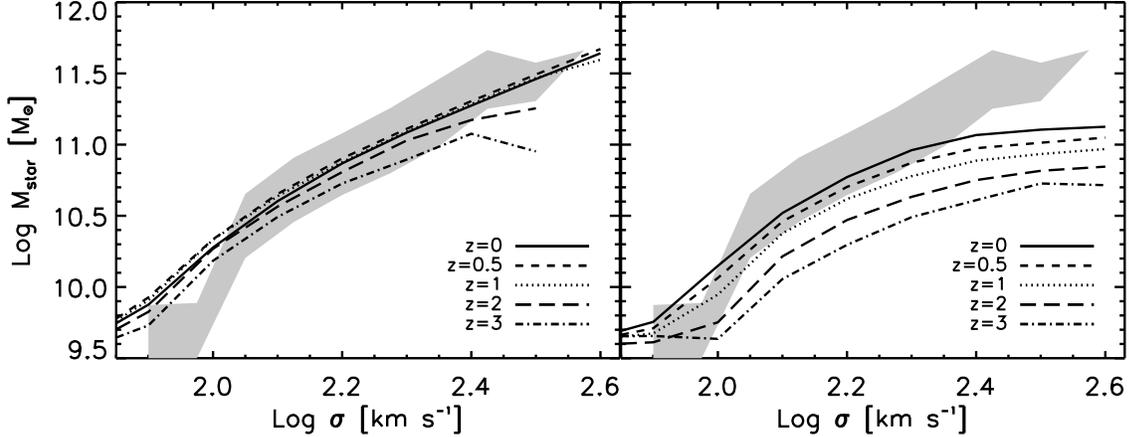}
    \caption{\emph{Left}: Predicted \mstar-\sis\ relation
    at different redshifts, as labelled,
    for a model a dark matter-dependent \sis\ as suggested by numerical
    simulations (\emph{left} panel) and a model
    with no dependence on dark matter (\emph{right} panel).
    Only galaxies with $B/T \gtrsim 0.9$ are considered.
    The predictions in both panels are compared to the subset of
    SDSS galaxies with same \bt\ cut (\emph{grey areas}).}
    \label{fig|MstarSigma}
\end{figure*}

\subsection{Additional constraints from velocity dispersions}
\label{subsec|Sigmas}

\begin{figure}
    \includegraphics[width=8.7truecm]{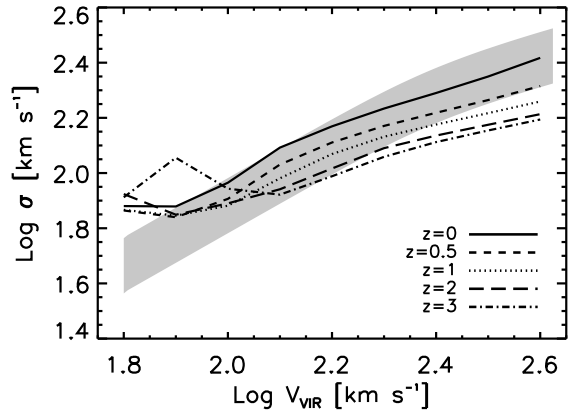}
    \caption{Predicted median \sis\ - \vvir\ relation at different redshifts, as labelled,
    for a model with a dark matter mass-dependent \sis\ and galaxies
    with $B/T>0.9$. The \emph{grey stripe} indicates the velocity dispersion-circular velocity
    correlation by \citet{Baes03} for early-type galaxies with circular velocity
    converted to velocity at the virial radius using the velocity-dependent correction of \citet{Dutton10a}.}
    \label{fig|SigmaVvir}
\end{figure}

Several additional clues on the hierarchical evolution
of spheroidal galaxies can be obtained when considering
velocity dispersions and the virial mass of galaxies.
In this section we will discuss the physical implications
that can be derived from the comparison of model predictions
with a variety of key observables that include velocity dispersion.
To this purpose,
unless otherwise stated, we will only consider here the subsample
of pure elliptical galaxies, i.e., those with $B/T \ge 0.9$ for which
a clear definition of velocity dispersion is possible and less biased,
both observationally and theoretically, given the near absence of
a stellar disc (although we note that a less extreme cut in \bt\ does not
alter the conclusions below).

All 3D velocity dispersions are computed following Eq.~(\ref{eq|CovingtonSigma}) and then converted,
consistently with what we discussed for sizes (Section~\ref{subsec|closercomparison}),
to line-of-sight 1D $\sigma(1D)$ using the \citet{Prugniel97} $S_K(n)$ coefficients,
i.e., $\sigma(1D) = [3S_K(n)]^{1/2} \sigma(3D)$,
where we set $n=4$ (the effect of the latter correction
is however relatively small and does not minimally alter our conclusions).

The very first correlation with velocity dispersion that is usually studied
is the one between luminosity/stellar mass and velocity dispersion, the Faber-Jackson relation \citep{FJ}.
Our results are presented in Figure~\ref{fig|MstarSigma} where the predicted \mstar-\sis\ relation is plotted
at the redshifts $z=0, 0.5, 1, 2, 3$ with a solid, dashed, dotted, long-dashed,
and dot-dashed line, respectively, as labelled. The grey band is the measured
\mstar-\sis\ relation from our subsample of SDSS galaxies with $B/T>0.9$.
The left panel of Figure~\ref{fig|MstarSigma} shows predictions
for a model with \sis\ computed via Eq.~(\ref{eq|CovingtonSigma}),
that includes dependence on the host dark matter halo, while in the right panel
\sis\ simply scales inversely with half-mass radius, $\propto \mstare/R_H$.
It is clear that the model with \sis\ computed via Eq.~(\ref{eq|CovingtonSigma})
provides a much better match to the data, with stellar mass properly increasing with increasing \sis.
Neglecting
any mass dependence in \sis\ inevitably produces a flattening at high masses where the model
predicts a quasi-linear dependence between half-mass radius and stellar mass
(cfr., e.g., Figure~\ref{fig|ReMsCompareWithData}). It is also interesting to note that neglecting any halo mass
dependence in \sis\ produces a much stronger evolution in velocity dispersion at fixed stellar mass, while a model
with halo mass dependence contains the evolution in \sis\ to $\lesssim 30\%$ in good agreement with direct observations
\citep[e.g.,][]{Bernardi09,Cenarro09,Cappellari09,Sande11}.

Several groups have proven that
a clear correlation exists between velocity dispersion and circular velocity
at the outer optical radius \citep[e.g.,][]{Ferrarese02,Baes03,Pizzella05,Chae11}.
This is expected from basic dark matter theory
\citep[e.g.,][]{Loeb03,Ciras05,Shankar06,Lapi09a} as during the
early fast-collapse phase of a halo, its potential well is established
and a dynamical link between baryon velocity at the center
and halo circular velocity should be established. Several high-resolution simulations have confirmed this behaviour
\citep[e.g.,][]{Zhao03,Diemand07,Wang11}.
We plots in Figure~\ref{fig|SigmaVvir} the predicted median correlation between
1D velocity dispersion and virial velocity \vvir\ of the halo at in fall \citep[see][]{Guo11},
at different redshifts, as labelled,
for the model with a dark matter mass-dependent \sis\ and galaxies with $B/T>0.9$.
The grey stripes in Figure~\ref{fig|SigmaVvir} indicate the correlation between
velocity dispersion and circular velocity inferred by \citet{Baes03}
for early-type galaxies with circular velocity converted to
velocity at the virial radius using the empirically-derived velocity-dependent
correction of \citet{Dutton10a} for early-type galaxies (their Eq. 3).
A good agreement is found in the local Universe. The model then predicts some
evolution in the zero point of this relation with galaxies at fixed velocity dispersion
being mapped into haloes with higher virial velocity but the
correlation is preserved despite mergers
\citep[e.g.,][]{Boylan05,Robertson06,Ciotti07}. We also find (not shown in the Figure)
that the scatter in velocity dispersion at fixed \vvir\ increases with increasing redshift,
in line with some observations \citep[][]{Courteau07,Ho07}.

\section{Discussion}
\label{sec|discu}

\subsection{Evolutionary features}
\label{subsec|evolution}

Having described here and in the previous sections
global structural properties of spheroids
of different masses and \bt, we now attempt to sketch a more
comprehensive picture of spheroid evolution. To this purpose,
we select subsamples of 100 galaxies of a given stellar mass and
\bt\ at $z=0$ and trace back in time the most massive progenitor
of each galaxy and record its properties.

\begin{figure}
    \centering
    \vspace{0.5cm}
    \includegraphics[width=7truecm]{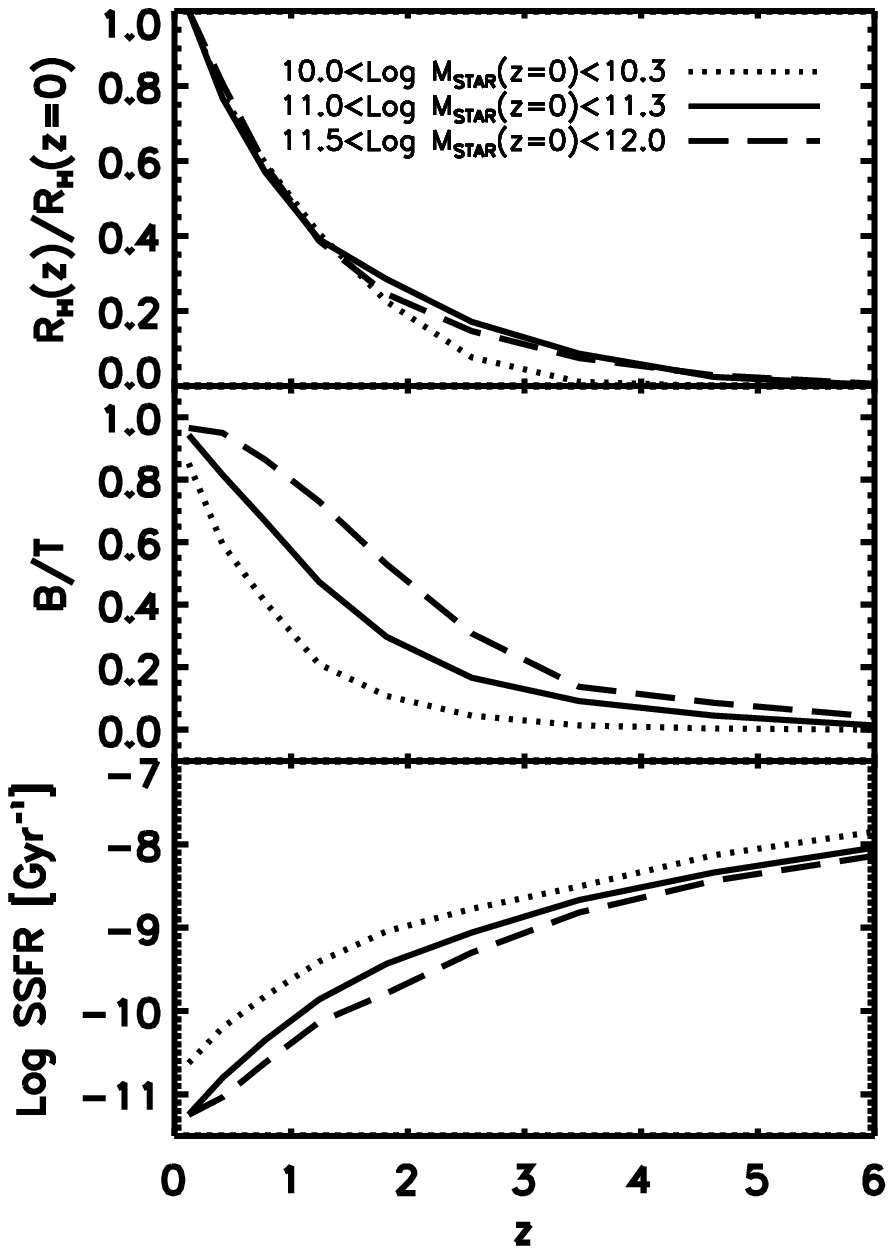}
    \caption{Evolution with redshift of the average properties of a
    subset of 100 galaxies derived by following the most massive progenitor
    back in time along its merger tree. We show results for three subsets
    of galaxies classified by having at $z=0$ $10<\log \mstare < 10.3$ (\emph{long-dashed} lines),
    $11<\log \mstare < 11.3$ (\emph{solid} lines), and $\log \mstare > 11.5$ (\emph{dotted} lines).
    All galaxies have $B/T>0.7$ at $z=0$. The \emph{top panel} shows the median fractional size evolution,
    the \emph{middle panel} the average bulge-to-total ratios, the \emph{bottom panel} the specific star formation rate.}
    \label{fig|Evolution}
\end{figure}

The result is given in Figure~\ref{fig|Evolution} which shows
the expected median redshift evolution of several properties
characterizing spheroid progenitors. We show results
for three subset of galaxies classified at
$z=0$ by having $10<\log \mstare < 10.3$ (long-dashed lines),
$11<\log \mstare < 11.3$ (solid lines), and $\log \mstare > 11.5$ (dotted lines).
All galaxies have $B/T>0.7$ at $z=0$.
The model predicts that the progenitors' half-mass radiuses
shrink when moving to higher redshifts (top panel; note that
we are here plotting progenitors identified from the merger trees,
while before we always considered \emph{different} galaxies
of the same stellar mass at different epochs).
Noticeably, all galaxies that end up
being large spheroids in the local Universe are predicted to share, on average,
quite similar size evolutions, at least at $z\lesssim 1.5-2$,
though different morphologies at higher redshifts.

\begin{figure*}
    \includegraphics[width=15truecm]{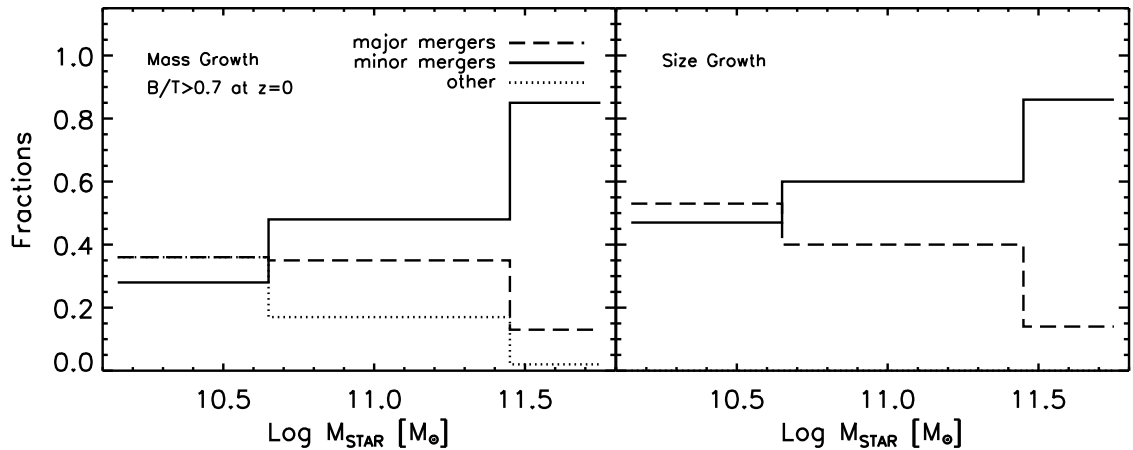}
    \caption{Fractional average mass (\emph{left}) and size (\emph{right}) growth of the
    subsets of galaxies discussed in Figure~\ref{fig|Evolution}.
    \emph{Long-dashed}, \emph{solid}, and \emph{dotted} lines refer,
    respectively, to the fractional cumulative growth down to $z=0$ as a
    function of final stellar mass, experienced by galaxies via major mergers,
    minor mergers, or any other mechanism (disc instability, star formation, etc...).}
    \label{fig|EvolutionFractions}
\end{figure*}

We find that most of the progenitors of bulge-dominated galaxies
at high redshifts are disc-dominated (middle panel). More specifically, galaxies that today lie
below the characteristic mass of $10^{11}\, \msune$, are found to turn
into disc-dominated systems at $z\sim 1.5-2$, with $B/T \lesssim 0.2$,
and extremely gas-rich (Figure~\ref{fig|ReMsWithDissipation}).
Even more massive galaxies with stellar mass at $z=0$ in the range
$10^{11}\, <\mstare/\msune \lesssim 2\times 10^{11}\, \msune$
rapidly turn into discs with a median $B/T \sim 0.3$.
In fact, we found that only 20\% of the galaxies
in this mass regime remain bulge-dominated
with $B/T \gtrsim 0.7$ at higher redshifts, with the majority
turning into disc systems with $B/T \sim 0.5$ by $z\gtrsim 1.5-2$.

The model also predicts that at all redshifts the most massive galaxies
are the least starforming galaxies, up to an order of magnitude at $z\lesssim 2$
(bottom panel).
The model therefore suggests that the observed
high redshift gas-rich, starforming, and clumpy
discs \citep[e.g.,][]{Forster11}, may be good candidates for being progenitors of today's
intermediate massive, early-type galaxies with $\mstare(z=0) \lesssim 10^{11}\, \msune$,
while the compact and red galaxies should end up being the most massive ellipticals we observe in the
local Universe. Full exploration of high and low number densities
of galaxies of a given property
will help to further constrain the model (see Section~\ref{subsec|additionalconstraints}).

We also note that beyond $z\gtrsim 2$ galaxies are observed to show a
flattening of their SSFR \citep[e.g.,][]{Gonzalez11},
while the predicted SSFR still continues to raise
beyond $z=2$ (bottom panel). Solving the discrepancy
between model and data is beyond the scope of the present paper.
We note however that it is a common feature of many galaxy formation models,
and it has been recently pointed out and discussed by
\citet{KhochfarSilk11} and \citet{Weinmann11}.

Figure~\ref{fig|EvolutionFractions} shows the fractional average mass (left) and size
(right) growth of the subsets of galaxies discussed in Figure~\ref{fig|Evolution}.
Long-dashed, solid, and dotted lines refer, respectively, to the fractional cumulative
growth down to $z=0$ as a function of final stellar mass, experienced by galaxies
via major mergers, minor mergers, or any other mechanism (such as disc instability
and/or in loco star formation). We find that the massive spheroids considered here
mainly grow through mergers in the model. Above
$\mstare \gtrsim 10^{11}\, \msune$ these galaxies grow more than $\gtrsim 50\%$ of
their final stellar mass and size via minor mergers,
while major mergers dominate the growth at lower masses
and become  progressively less important at higher masses \citep[e.g.,][]{Khochfar09a}.
This is expected given that the median accretion history of their
typical host dark matter haloes, in the range $10^{12}-10^{13}\,
h^{-1}\, $\msun, are found in
high-resolution numerical simulations to be dominated by mergers with satellites
that are $\sim$ 10\% of the final halo mass \citep[e.g.,][]{Stewart08}.

We also find that most of the minor mergers, especially in the most massive galaxies, are dry, i.e., have a (cold) gas mass fraction in the progenitors
that is lower than 0.15. Minor dry mergers can
roughly preserve the projections of the fundamental plane \citep[e.g.,][and references therein]{Ciotti09},
and galactic central densities \citep[e.g.,][and references therein]{Cimatti08,Bezanson09},
though they may also increase the scatter in the scaling relations
\citep[e.g.,][and references therein]{Nipoti09}.
The findings of Figure~\ref{fig|EvolutionFractions} is consistent with an inside-out evolutionary scenario, where
stellar matter is continuously added to the outskirts of the compact
high-redshift galaxies as time goes on \citep[e.g.,][]{Naab09,Oser10}.
More detailed comparisons with metallicity, age and colour gradients
are needed to set this model on firmer foots (see discussion in Section~\ref{subsec|additionalconstraints}).

\subsection{Comparison with other models}
\label{subsec|othermodelscompare}

While all galaxy formation models have discussed predictions
for stellar mass distributions, just a handful have taken a step
further to also consider structural properties, especially for early-type galaxies.

\citet{KochfarSilk06Rez}
within the context of a full semi-analytic model,
emphasized the role of gas dissipation in mergers
in forming compact massive spheroids at high redshifts,
deriving progenitors' gas fractions in good agreement with the ones
discussed here.

\citet{Almeida07,Almeida08}, following \citet{Cole00},
studied the galaxy size-mass and global fundamental plane relations
predicted by two significantly different renditions of the GALFORM
semi-analytic model by \citet{Baugh05} and \citet{Bower06}.
After varying most of the parameters relevant for determining bulge sizes,
they concluded that \emph{both} models fail to reproduce the sizes of bright
early-type galaxies, though they noted that
a better match to the data was achieved with no adiabatic
contraction \citep[e.g.,][]{Tissera10,Covington11},
as assumed in this work.
On a similar note, \citet{Gonzalez09}, following on \citet{Almeida07},
further studied the galaxy size-mass relations in the GALFORM models
varying other parameters, such as orbital energy in the merger,
but still finding significant disagreement with the data.

\citet{ShankarRe} and \citet{ShankarPhire} showed that
the size-mass relation at $z=0$ predicted by the Bower et al.
model is much flatter than the observed one due to too large
low-mass galaxies with stellar mass $\mstare \lesssim 10^{11}\, \msune$,
similarly to what found here. They suggested that the latter problem
may be linked to the initial conditions, given that large and low-mass
galaxies are present at all epochs in the model, in line
with what discussed here.

We have also adapted the recipes for bulge size growth
via mergers and disc instability discussed in Section~\ref{subsec|RefModel} to a previous
version of the Munich code by \citet{DeLucia07}. The latter
differs from the \citet{Guo11} version in several respects,
from physical recipes to values of the best-fit parameters.
We have checked, however, that most of the results
discussed here remain globally similar.

\citet{Monaco07} implemented the recipes from
\citet{Cole00} for size growth
in their MORGANA model (with no dissipation) finding acceptable agreement with the local
size-mass relation for early-type galaxies, though with a
scatter larger than the observed one (see their Figure 16).
Understanding the success of the latter model with respect to the previous mentioned ones
relies on pinning down the differences in the physical inputs of the
MORGANA model and the impact they have on size evolution.

Other works aimed at studying the global structural properties
of local early-type galaxies through semi-analytic techniques
was pursued by \citet{Ciras05}, within the framework of the
\citet{Granato04} model for the co-evolution of super-massive
black holes and their host massive spheroids. Cirasuolo et al.
showed that by tightening velocity dispersion of spheroids
to the virial velocity at the epoch of their formation \citep[see also][]{Loeb03},
the local early-type velocity dispersion function
\citep{sheth03,Shankar04,Bernardi10} and Gaussian dispersion
in sizes was fully recovered. Their study is quite
intriguing as the match to the photometric and dynamical
properties of local ellipticals only relies on galaxy properties at
virialization epoch and minimizes the role of later merger events.

A more refined theory of structural evolution
of spheroids besides mergers, has been presented by \citep{Fan08,Fan10}.
As anticipated in Section~\ref{sec|intro}, this class of models explains size
evolution of spheroids via expansion consequent to the blow-out of
substantial amounts of mass via quasar and/or stellar feedback
\citep{Fan08,Fan10,Damjanov09}. Numerical support to the latter models
was recently provided by \citet{RagoneGranato} who showed that even
in the presence of dark matter, baryons can indeed expand by a factor
of a few consequent to substantial mass losses. They also pointed out that the puffing up
via expansion may be too rapid with respect to the old ages
measured for the compact high-z early-type galaxies. Understanding the actual
role played by expansion versus mergers is beyond the scope of the present work.
However, we discussed in Section~\ref{subsec|evolredshift}
that the strong size evolution for the most massive ellipticals
claimed by some observational groups \citep{Ryan10,Huertas12,Newman12} can hardly be reproduced
in our only-merger model, and possibly some extra-expansion (a factor of $\sim 2$)
in the first phases of evolution might help.

Relevant numerical work has been pursued in the
last years to explore the size evolution of spheroids in a
full cosmological context. We recall here the work by \citet{Naab09}
and \citet[][see also Scannapieco et al. 2011]{Oser10},
who developed high resolution hydrodynamical cosmological simulations
of massive spheroidal galaxies. They particularly emphasized
that galaxies above the characteristic mass of
$\mstare \gtrsim 10^{11}\, \msune$ can accrete via minor
merger about 80\% of their final stellar mass, in
agreement with what found here (Figure~\ref{fig|EvolutionFractions}).

\subsection{General issues and additional constraints}
\label{subsec|additionalconstraints}

In this work we emphasized the role of gas dissipation
as a viable mechanism to shrink bulge sizes, especially in lower mass
galaxies, thus improving the match to a variety of different observables.
Gas dissipation is indeed a natural outcome of gas-rich mergers and thus should
to be properly included in complete models of galaxy formation.
Nevertheless, we cannot rule out that part of the discrepancy between
model predictions and data, especially regarding the match with the size mass relation
(Section~\ref{subsec|includingdissipation}), could also be induced by some other wrong model inputs.
For example, it has been recognized that this SAM, like several others, overproduces
the stellar mass function at high redshifts and low stellar masses \citep[e.g.,][]{Henriquez12}. This in turn
implies more massive and larger galaxies in lower mass haloes with
respect to what expected from, e.g., cumulative abundance matching arguments
\citep[][]{Moster12}, thus possibly contributing
to the flattening at low masses in the size mass relation (Section~\ref{subsec|firstcomparison}).
Besides the actual performance of the model in properly predicting the size-mass relation,
we nevertheless stress that gas dissipation can improve the match to several other observables,
as discussed above and further in the Appendices (e.g., size-age relation, correlations with velocity dispersion, etc...). 

In this work we have focused our attention on the most relevant
median scaling relations among structural properties of bulged galaxies.
It is clear, however, that several other properties are equally important
to pin down and better constrain the viable models.

First of all, it is fundamental for a model not only
to produce the correct structure but also the correct number of galaxies of a given type.
Galaxy formation models have most seriously investigated the match to the stellar mass function,
finding good agreement adopting different physical prescriptions \citep[e.g.,][]{Cole00,Benson03,Granato04,Croton06},
though significant uncertainties still affect high and low redshift measurements
\citep{Bernardi10,Marchesini09,Marchesini10}.

The study of number densities in other properties such as size,
is instead still at its infancy. The first study was carried out by \citet{Cole00}.
A more recent attempt has been carried out by \citet{ShankarPhire} who compared detailed
predictions from hierarchical models with the latest
releases of the $\Phi(R_e)$ galaxy size function, finding
that the \citet{Bower06} model overpredicts the number of very compact and very large galaxies
\citep[see also][]{Trujillo09,Taylor10}.

Higher redshift measurements of portions of the size
function of spheroids are also now becoming available
\citep[e.g.,][]{Mancini09,Saracco10,Valentinuzzi10a,VanDokkum10}.
At higher redshifts the scatter in sizes at fixed stellar
mass apparently seems to increase as large and compact galaxies
seem to co-exist at the same epoch \citep[e.g.,][]{Mancini10,Saracco11}.
This increase in scatter may suggest a faster evolution in sizes at fixed
stellar mass \citep{Fan10}, although trends between size and age/star formation rate, may bias this result
\citep[e.g.,][]{Mosleh11,Forster11,Saracco11}.

In order to use number density measurements of size
and velocity functions \citep{Ciras05} at different redshifts,
a homogeneous and well studied spectroscopic and photometric
sample of galaxies is needed, now not yet released.
Larger statistical samples will become available in the near future
(such as multi-wavelength optical surveys from the Canada French Hawaii Telescope
and the Next Generation Virgo Cluster Survey).

Ages, colours, metallicities, and other dynamical properties \citep[e.g.,][]{Cappellari09}
can be key observables to probe galaxy evolution.
For example, \citet{Bernardi11a} and \citet{Bernardi11b}, making use of
the latest SDSS data releases, showed that all correlations with
stellar mass steepen above $\mstare \gtrsim 2\times 10^{11} \msune$
\citep[see also][]{Fasano10,vanderwel11},
while relations with velocity dispersion don't.
Bernardi et al. claimed, in line with other
observational works \citep[e.g.,][]{Kauff04,vanderwel09} and with the results
presented in Figure~\ref{fig|EvolutionFractions},
the presence of two mass scales, one at $\mstare \sim 3\times 10^{10}\, \msune$, below
which gas dissipation controls galaxy formation, and a higher mass scale
$\mstare \gtrsim 2\times 10^{11}\, \msune$ above which mergers dominate the evolution.
The empirical mass scales
noticed by Bernardi et al. are in line with the
characteristic masses $M_c$ and $M_s$ emphasized in Section~\ref{subsec|firstcomparison}.
Metallicity gradients have been recently calibrated by a number of groups
\citep[e.g.,][]{Foster09,Spolaor10b,Forbes11} and can provide invaluable
insights into the evolutionary patterns of early-type galaxies
\citep[e.g.,][]{DiMatteo09,LaBarbera09}.

\section{Conclusions}
\label{sec|conclu}

Understanding structural evolution of spheroids has become one of the hottest topics in cosmology
in the last years as it can provide invaluable insights into the true physical processes
that regulated galaxy evolution. While angular momentum conservation may explain many properties of discs,
the origin of bulges is still largely debated.
The situation is even more puzzling given that at higher redshifts galaxies
present further disparate structural and physical properties, from clumpy star-forming discs,
to very compact, red and massive galaxies. Such a complicated zoology is difficult to reconcile within
a coherent framework of galaxy formation, and in fact we discussed in Section~\ref{sec|intro}
that models sometimes propose conflicting scenarios.

Our aim in this work was to study the predictions of a state-of-the-art
hierarchical model of galaxy formation, which
evolves the sizes of spheroids via mergers and disc instabilities,
against the most recent local and high redshift data.
To this purpose we updated the source code of the latest release of the
Munich semi-analytic galaxy formation model by \citet{Guo11},
by modifying the computation of bulge radii exploring a variety
of possibilities, and added the calculation of the coupled velocity dispersions.
In order to properly compare model predictions with available data
we made use of a large sample of early-type galaxies from SDSS for which bulge-to-disc
decompositions have been performed both via the SDSS automated
``on-the-fly'' analysis and by applying a detailed fitting code.

Our main results can be summarized as follows.
\begin{itemize}
  \item Sizes are computed in the model via energy conservation in dissipationless mergers.
   Taking the model at face value
   at masses below $M_s \lesssim 10^{11}\, \msune$ the model predicts a flattening of the size-mass
   relation at variance with the data, already at $z\sim 2$.
  \item Following the results of hydro-simulations, we have included the energy dissipated in gas-rich
  \emph{major} mergers in the energy budget. This modification produces progressively more compact
 remnants with decreasing stellar mass, proportionally to the fraction of cold gas in the progenitors,
  improving the match with a variety of observables.
     \item  We confirm and discuss evidence for two characteristic masses.
   One is at $M_c \sim 3 \times 10^{10}\, \msune$, below which the initial bulge sizes are
   controlled by dissipation (higher gas fractions) and then evolve under mergers and
   disc instabilities. Galaxies above $M_s \gtrsim 10^{11}\, \msune$, instead
   mainly grow through minor dry mergers, especially at $z<1$ (Figure~\ref{fig|EvolutionFractions}).
  \item We find that the global scatter (1-$\sigma$ uncertainty)
in sizes at fixed stellar mass for galaxies with $B/T>0.7$ is comparable, though systematically higher by $\lesssim 40\%$
than the observed one. The predicted amount of scatter for galaxies
with stellar mass $\mstare \gtrsim 3\times 10^{10}\, \msune$ does not
depend much on dissipation and/or amount of orbital energy in the merger.
\item Spheroids are predicted to be, on average, more compact at higher
  redshifts at fixed stellar mass. More specifically, at fixed \bt\
  a nearly mass-independent relatively mild decrease in sizes is predicted,
  in possible disagreement with some observations.
  \item The model predicts that environment plays a significant role in defining the
  structural properties of bulged galaxies.
  (Central) galaxies residing in denser environments are predicted to undergo
  more mergers, thus evolve faster and end up having larger sizes at fixed stellar mass.
  \item The progenitors of the massive spheroids with $\mstare \sim (1-2) \times 10^{11}\, \msune$ today
  with $B/T>0.7$ are predicted to be compact,
  low starforming $z \sim 2$ protogalaxies with a median $B/T \sim 0.3$, with only $\sim 20\%$
  remaining bulge-dominated.
  The progenitors of lower-mass spheroids with $B/T>0.7$ tend to be closer to the starforming and gas-rich
  proto-discs observed at similar redshifts.
   \item Finally, we also discuss a number of related issues (in the text and in the Appendices), ranging from
   the correlations with galaxy age,
   with central black hole mass, with velocity dispersion, to the scaling relations of pseudobulges.
\end{itemize}

\section*{Acknowledgments}
FS warmly thanks Enzo Gattorno for his kind and helpful support.
We thank Bruno Henriquez and Gerard Lemson for many useful inputs.
We thank Guinevere Kauffmann and Simon White for making the source code of the
Munich model available to us to properly carry out this work and for general comments.
We also thank Qi Guo, Luca Ciotti, Luca Graziani, Sadegh Khochfar,
Juan Gonz\'{a}lez, Frederic Bournaud, Simone Weinmann,
Ravi Sheth, Luigi Danese, Andrea Lapi, Alister Graham, Eyal Neistein,
Fabio Fontanot, Silvia Bonoli, Thorsten Naab, Avishai Dekel, and Matt Covington for several interesting discussions.
We thank Dimitri Gadotti and David Fisher for providing us their data on pseudobulges.
FS acknowledges support from the Alexander von Humboldt Foundation and
partial support from a Marie Curie grant.
MB is grateful for support provided by NASA grant ADP/NNX09AD02G.
We thank the referee for a useful report that significantly
improved the presentation of the paper.

\bibliographystyle{mn2e}
\bibliography{../RefMajor}

\begin{thebibliography}{}

\bibitem[\protect\citeauthoryear{{Almeida}, {Baugh} \& {Lacey}}{{Almeida}
  et~al.}{2007}]{Almeida07}
{Almeida} C.,  {Baugh} C.~M.,    {Lacey} C.~G.,  2007, \mnras, 376, 1711

\bibitem[\protect\citeauthoryear{{Almeida}, {Baugh}, {Wake}, {Lacey}, {Benson},
  {Bower} \& {Pimbblet}}{{Almeida} et~al.}{2008}]{Almeida08}
{Almeida} C.,  {Baugh} C.~M.,  {Wake} D.~A.,  {Lacey} C.~G.,  {Benson} A.~J.,
  {Bower} R.~G.,    {Pimbblet} K.,  2008, \mnras, 386, 2145

\bibitem[\protect\citeauthoryear{{Ascaso}, {Aguerri}, {Varela}, {Cava},
  {Bettoni}, {Moles} \& {D'Onofrio}}{{Ascaso} et~al.}{2011}]{Ascaso11}
{Ascaso} B.,  {Aguerri} J.~A.~L.,  {Varela} J.,  {Cava} A.,  {Bettoni} D.,
  {Moles} M.,    {D'Onofrio} M.,  2011, \apj, 726, 69

\bibitem[\protect\citeauthoryear{{Auger}, {Treu}, {Brewer} \&
  {Marshall}}{{Auger} et~al.}{2011}]{Auger11}
{Auger} M.~W.,  {Treu} T.,  {Brewer} B.~J.,    {Marshall} P.~J.,  2011, \mnras,
  411, L6

\bibitem[\protect\citeauthoryear{{Baes}, {Buyle}, {Hau} \& {Dejonghe}}{{Baes}
  et~al.}{2003}]{Baes03}
{Baes} M.,  {Buyle} P.,  {Hau} G.~K.~T.,    {Dejonghe} H.,  2003, \mnras, 341,
  L44

\bibitem[\protect\citeauthoryear{{Barnes}}{{Barnes}}{1992}]{Barnes92}
{Barnes} J.~E.,  1992, \apj, 393, 484

\bibitem[\protect\citeauthoryear{{Baugh}, {Lacey}, {Frenk}, {Granato}, {Silva},
  {Bressan}, {Benson} \& {Cole}}{{Baugh} et~al.}{2005}]{Baugh05}
{Baugh} C.~M.,  {Lacey} C.~G.,  {Frenk} C.~S.,  {Granato} G.~L.,  {Silva} L.,
  {Bressan} A.,  {Benson} A.~J.,    {Cole} S.,  2005, \mnras, 356, 1191

\bibitem[\protect\citeauthoryear{{Bennert}, {Auger}, {Treu}, {Woo} \&
  {Malkan}}{{Bennert} et~al.}{2011}]{Bennert11}
{Bennert} V.~N.,  {Auger} M.~W.,  {Treu} T.,  {Woo} J.,    {Malkan} M.~A.,
  2011, \apj, 708, 1507

\bibitem[\protect\citeauthoryear{{Benson}, {Frenk}, {Baugh}, {Cole} \&
  {Lacey}}{{Benson} et~al.}{2003}]{Benson03}
{Benson} A.~J.,  {Frenk} C.~S.,  {Baugh} C.~M.,  {Cole} S.,    {Lacey} C.~G.,
  2003, \mnras, 343, 679

\bibitem[\protect\citeauthoryear{{Bernardi}}{{Bernardi}}{2009}]{Bernardi09}
{Bernardi} M.,  2009, \mnras, 395, 1491

\bibitem[\protect\citeauthoryear{{Bernardi}, {Shankar}, {Hyde}, {Mei},
  {Marulli} \& {Sheth}}{{Bernardi} et~al.}{2010}]{Bernardi10}
{Bernardi} M.,  {Shankar} F.,  {Hyde} J.~B.,  {Mei} S.,  {Marulli} F.,
  {Sheth} R.~K.,  2010, \mnras, 404, 2087

\bibitem[\protect\citeauthoryear{{Bernardi}, {Roche}, {Shankar} \&
  {Sheth}}{{Bernardi} et~al.}{2011a}]{Bernardi11a}
{Bernardi} M.,  {Roche} N.,  {Shankar} F.,    {Sheth} R.~K.,  2011a, \mnras,
  412, 684

\bibitem[\protect\citeauthoryear{{Bernardi}, {Roche}, {Shankar} \&
  {Sheth}}{{Bernardi} et~al.}{2011b}]{Bernardi11b}
{Bernardi} M.,  {Roche} N.,  {Shankar} F.,    {Sheth} R.~K.,  2011b, \mnras,
  412, L6

\bibitem[\protect\citeauthoryear{{Bezanson}, {van Dokkum}, {Tal}, {Marchesini},
  {Kriek}, {Franx} \& {Coppi}}{{Bezanson} et~al.}{2009}]{Bezanson09}
{Bezanson} R.,  {van Dokkum} P.~G.,  {Tal} T.,  {Marchesini} D.,  {Kriek} M.,
  {Franx} M.,    {Coppi} P.,  2009, \apj, 697, 1290

\bibitem[\protect\citeauthoryear{{Blain}, {Smail}, {Ivison}, {Kneib} \&
  {Frayer}}{{Blain} et~al.}{2002}]{Blain02}
{Blain} A.~W.,  {Smail} I.,  {Ivison} R.~J.,  {Kneib} J.-P.,    {Frayer} D.~T.,
   2002, \physrep, 369, 111

\bibitem[\protect\citeauthoryear{{Bonoli}, {Marulli}, {Springel}, {White},
  {Branchini} \& {Moscardini}}{{Bonoli} et~al.}{2009}]{Bonoli09}
{Bonoli} S.,  {Marulli} F.,  {Springel} V.,  {White} S.~D.~M.,  {Branchini} E.,
     {Moscardini} L.,  2009, \mnras, 396, 423

\bibitem[\protect\citeauthoryear{{Bonoli}, {Shankar}, {White}, {Springel} \&
  {Wyithe}}{{Bonoli} et~al.}{2010}]{Bonoli10}
{Bonoli} S.,  {Shankar} F.,  {White} S.~D.~M.,  {Springel} V.,    {Wyithe}
  J.~S.~B.,  2010, \mnras, 404, 399

\bibitem[\protect\citeauthoryear{{Bournaud}, {Chapon}, {Teyssier}, {Powell},
  {Elmegreen}, {Elmegreen}, {Duc}, {Contini}, {Epinat} \& {Shapiro}}{{Bournaud}
  et~al.}{2011a}]{Bournaud11a}
{Bournaud} F.,  {Chapon} D.,  {Teyssier} R.,  {Powell} L.~C.,  {Elmegreen}
  B.~G.,  {Elmegreen} D.~M.,  {Duc} P.,  {Contini} T.,  {Epinat} B.,
  {Shapiro} K.~L.,  2011a, \apj, 730, 4

\bibitem[\protect\citeauthoryear{{Bournaud}, {Dekel}, {Teyssier}, {Cacciato},
  {Daddi}, {Juneau} \& {Shankar}}{{Bournaud} et~al.}{2011b}]{Bournaud11b}
{Bournaud} F.,  {Dekel} A.,  {Teyssier} R.,  {Cacciato} M.,  {Daddi} E.,
  {Juneau} S.,    {Shankar} F.,  2011b, \apjl, 741, L33

\bibitem[\protect\citeauthoryear{{Bower}, {Benson}, {Malbon}, {Helly}, {Frenk},
  {Baugh}, {Cole} \& {Lacey}}{{Bower} et~al.}{2006}]{Bower06}
{Bower} R.~G.,  {Benson} A.~J.,  {Malbon} R.,  {Helly} J.~C.,  {Frenk} C.~S.,
  {Baugh} C.~M.,  {Cole} S.,    {Lacey} C.~G.,  2006, \mnras, 370, 645

\bibitem[\protect\citeauthoryear{{Boylan-Kolchin}, {Ma} \&
  {Quataert}}{{Boylan-Kolchin} et~al.}{2005}]{Boylan05}
{Boylan-Kolchin} M.,  {Ma} C.-P.,    {Quataert} E.,  2005, \mnras, 362, 184

\bibitem[\protect\citeauthoryear{{Boylan-Kolchin}, {Springel}, {White},
  {Jenkins} \& {Lemson}}{{Boylan-Kolchin} et~al.}{2009}]{Boylan09}
{Boylan-Kolchin} M.,  {Springel} V.,  {White} S.~D.~M.,  {Jenkins} A.,
  {Lemson} G.,  2009, \mnras, 398, 1150

\bibitem[\protect\citeauthoryear{{Buitrago}, {Trujillo}, {Conselice},
  {Bouwens}, {Dickinson} \& {Yan}}{{Buitrago} et~al.}{2008}]{Buitrago08b}
{Buitrago} F.,  {Trujillo} I.,  {Conselice} C.~J.,  {Bouwens} R.~J.,
  {Dickinson} M.,    {Yan} H.,  2008, \apjl, 687, L61

\bibitem[\protect\citeauthoryear{{Bullock}, {Kolatt}, {Sigad}, {Somerville},
  {Kravtsov}, {Klypin}, {Primack} \& {Dekel}}{{Bullock}
  et~al.}{2001}]{Bullock01}
{Bullock} J.~S.,  {Kolatt} T.~S.,  {Sigad} Y.,  {Somerville} R.~S.,  {Kravtsov}
  A.~V.,  {Klypin} A.~A.,  {Primack} J.~R.,    {Dekel} A.,  2001, \mnras, 321,
  559

\bibitem[\protect\citeauthoryear{{Cappellari}, {di Serego Alighieri},
  {Cimatti}, {Daddi}, {Renzini}, {Kurk}, {Cassata}, {Dickinson},
  {Franceschini}, {Mignoli}, {Pozzetti}, {Rodighiero}, {Rosati} \&
  {Zamorani}}{{Cappellari} et~al.}{2009}]{Cappellari09}
{Cappellari} M. et al., 2009, \apjl, 704, L34

\bibitem[Catinella et al.(2010)]{Catinella10} Catinella B. et al., 2010, \mnras, 403, 683

\bibitem[\protect\citeauthoryear{{Cenarro} \& {Trujillo}}{{Cenarro} \&
  {Trujillo}}{2009}]{Cenarro09}
{Cenarro} A.~J.,  {Trujillo} I.,  2009, \apjl, 696, L43

\bibitem[Chabrier(2003)]{Chabrier03} Chabrier G., 2003, \pasp, 
115, 763 

\bibitem[\protect\citeauthoryear{{Chae}}{{Chae}}{2011}]{Chae11}
{Chae} K.-H.,  2011, \mnras, 413, 887

\bibitem[\protect\citeauthoryear{{Chapman}, {Neri}, {Bertoldi}, {Smail},
  {Greve}, {Trethewey}, {Blain}, {Cox}, {Genzel}, {Ivison}, {Kovacs}, {Omont}
  \& {Swinbank}}{{Chapman} et~al.}{2008}]{Chapman08}
{Chapman} S.~C. et al., 2008, \apj, 689, 889

\bibitem[\protect\citeauthoryear{{Cimatti}, {Cassata}, {Pozzetti}, {Kurk},
  {Mignoli}, {Renzini}, {Daddi}, {Bolzonella}, {Brusa}, {Rodighiero},
  {Dickinson}, {Franceschini}, {Zamorani}, {Berta}, {Rosati} \&
  {Halliday}}{{Cimatti} et~al.}{2008}]{Cimatti08}
{Cimatti} A. et al., 2008, \aap, 482, 21

\bibitem[\protect\citeauthoryear{{Cimatti}, {Nipoti} \& {Cassata}}{{Cimatti}
  et~al.}{2012}]{Cimatti12}
{Cimatti} A.,  {Nipoti} C.,    {Cassata} P.,  2012, \mnras, 422, L62

\bibitem[Ciotti 
\& van Albada(2001)]{Ciotti01} Ciotti L., van Albada T.~S., 2001, \apjl, 552, L13 

\bibitem[Ciotti et al.(2007)]{Ciotti07} Ciotti L., Lanzoni B., 
Volonteri M., 2007, \apj, 658, 65 

\bibitem[\protect\citeauthoryear{{Ciotti}}{{Ciotti}}{2009}]{CiottiReview}
{Ciotti} L.,  2009, Nuovo Cimento Rivista Serie, 32, 1

\bibitem[\protect\citeauthoryear{{Ciotti}, {Ostriker} \& {Proga}}{{Ciotti}
  et~al.}{2009}]{Ciotti09}
{Ciotti} L.,  {Ostriker} J.~P.,    {Proga} D.,  2009, ArXiv e-prints

\bibitem[\protect\citeauthoryear{{Cirasuolo}, {Shankar}, {Granato}, {De Zotti}
  \& {Danese}}{{Cirasuolo} et~al.}{2005}]{Ciras05}
{Cirasuolo} M.,  {Shankar} F.,  {Granato} G.~L.,  {De Zotti} G.,    {Danese}
  L.,  2005, \apj, 629, 816

\bibitem[\protect\citeauthoryear{{Cole}, {Lacey}, {Baugh} \& {Frenk}}{{Cole}
  et~al.}{2000}]{Cole00}
{Cole} S.,  {Lacey} C.~G.,  {Baugh} C.~M.,    {Frenk} C.~S.,  2000, \mnras,
  319, 168

\bibitem[\protect\citeauthoryear{{Cooper}, {Griffith}, {Newman}, {Coil},
  {Davis}, {Dutton}, {Faber}, {Guhathakurta}, {Koo}, {Lotz}, {Weiner},
  {Willmer} \& {Yan}}{{Cooper} et~al.}{2012}]{Cooper12}
{Cooper} M.~C. et al.,  2012, \mnras, 419,
  3018

\bibitem[\protect\citeauthoryear{{Courteau}, {McDonald}, {Widrow} \&
  {Holtzman}}{{Courteau} et~al.}{2007}]{Courteau07}
{Courteau} S.,  {McDonald} M.,  {Widrow} L.~M.,    {Holtzman} J.,  2007, \apjl,
  655, L21

\bibitem[\protect\citeauthoryear{{Covington}, {Dekel}, {Cox}, {Jonsson} \&
  {Primack}}{{Covington} et~al.}{2008}]{Covington08}
{Covington} M.,  {Dekel} A.,  {Cox} T.~J.,  {Jonsson} P.,    {Primack} J.~R.,
  2008, \mnras, 384, 94

\bibitem[\protect\citeauthoryear{{Covington}, {Primack}, {Porter}, {Croton},
  {Somerville} \& {Dekel}}{{Covington} et~al.}{2011}]{Covington11}
{Covington} M.~D.,  {Primack} J.~R.,  {Porter} L.~A.,  {Croton} D.~J.,
  {Somerville} R.~S.,    {Dekel} A.,  2011, \mnras, 415, 3135

\bibitem[\protect\citeauthoryear{{Croton}, {Springel}, {White}, {De Lucia},
  {Frenk}, {Gao}, {Jenkins}, {Kauffmann}, {Navarro} \& {Yoshida}}{{Croton}
  et~al.}{2006}]{Croton06}
{Croton} D.~J. et al., 2006, \mnras, 365, 11

\bibitem[Croton(2006)]{Croton06b} Croton D.~J., 2006, \mnras, 
369, 1808 

\bibitem[\protect\citeauthoryear{{Damjanov}, {McCarthy}, {Abraham},
  {Glazebrook}, {Yan}, {Mentuch}, {LeBorgne}, {Savaglio}, {Crampton},
  {Murowinski}, {Juneau}, {Carlberg}, {J{\o}rgensen}, {Roth}, {Chen} \&
  {Marzke}}{{Damjanov} et~al.}{2009}]{Damjanov09}
{Damjanov} I. et al., 2009, \apj, 695, 101

\bibitem[\protect\citeauthoryear{{De Lucia} \& {Blaizot}}{{De Lucia} \&
  {Blaizot}}{2007}]{DeLucia07}
{De Lucia} G.,  {Blaizot} J.,  2007, \mnras, 375, 2

\bibitem[\protect\citeauthoryear{{De Lucia}, {Boylan-Kolchin}, {Benson},
  {Fontanot} \& {Monaco}}{{De Lucia} et~al.}{2010}]{DeLucia10}
{De Lucia} G.,  {Boylan-Kolchin} M.,  {Benson} A.~J.,  {Fontanot} F.,
  {Monaco} P.,  2010, \mnras, 406, 1533

\bibitem[\protect\citeauthoryear{{De Lucia}, {Springel}, {White}, {Croton} \&
  {Kauffmann}}{{De Lucia} et~al.}{2006}]{DeLucia06}
{De Lucia} G.,  {Springel} V.,  {White} S.~D.~M.,  {Croton} D.,    {Kauffmann}
  G.,  2006, \mnras, 366, 499

\bibitem[\protect\citeauthoryear{{de Vaucouleurs}}{{de
  Vaucouleurs}}{1948}]{deVac}
{de Vaucouleurs} G.,  1948, Annales d'Astrophysique, 11, 247

\bibitem[\protect\citeauthoryear{{Decarli}, {Falomo}, {Treves}, {Labita},
  {Kotilainen} \& {Scarpa}}{{Decarli} et~al.}{2010}]{Decarli10}
{Decarli} R.,  {Falomo} R.,  {Treves} A.,  {Labita} M.,  {Kotilainen} J.~K.,
  {Scarpa} R.,  2010, \mnras, 402, 2453

\bibitem[\protect\citeauthoryear{{Dekel}, {Birnboim}, {Engel}, {Freundlich},
  {Goerdt}, {Mumcuoglu}, {Neistein}, {Pichon}, {Teyssier} \& {Zinger}}{{Dekel}
  et~al.}{2009a}]{Dekel09a}
{Dekel} A. et al., 2009a, \nat, 457, 451

\bibitem[\protect\citeauthoryear{{Dekel}, {Sari} \& {Ceverino}}{{Dekel}
  et~al.}{2009b}]{Dekel09b}
{Dekel} A.,  {Sari} R.,    {Ceverino} D.,  2009b, \apj, 703, 785

\bibitem[\protect\citeauthoryear{{Di Matteo}, {Jog}, {Lehnert}, {Combes} \&
  {Semelin}}{{Di Matteo} et~al.}{2009}]{DiMatteo09}
{Di Matteo} P.,  {Jog} C.~J.,  {Lehnert} M.~D.,  {Combes} F.,    {Semelin} B.,
  2009, \aap, 501, L9 

\bibitem[\protect\citeauthoryear{{Diemand}, {Kuhlen} \& {Madau}}{{Diemand}
  et~al.}{2007}]{Diemand07}
{Diemand} J.,  {Kuhlen} M.,    {Madau} P.,  2007, \apj, 667, 859

\bibitem[\protect\citeauthoryear{{Djorgovski} \& {Davis}}{{Djorgovski} \&
  {Davis}}{1987}]{DDFP}
{Djorgovski} S.,  {Davis} M.,  1987, \apj, 313, 59

\bibitem[\protect\citeauthoryear{{Dutton}, {Conroy}, {van den Bosch}, {Prada}
  \& {More}}{{Dutton} et~al.}{2010}]{Dutton10a}
{Dutton} A.~A.,  {Conroy} C.,  {van den Bosch} F.~C.,  {Prada} F.,    {More}
  S.,  2010, \mnras, 407, 2

\bibitem[\protect\citeauthoryear{{Eggen}, {Lynden-Bell} \& {Sandage}}{{Eggen}
  et~al.}{1962}]{Eggen62}
{Eggen} O.~J.,  {Lynden-Bell} D.,    {Sandage} A.~R.,  1962, \apj, 136, 748

\bibitem[\protect\citeauthoryear{{Erb}, {Steidel}, {Shapley}, {Pettini},
  {Reddy} \& {Adelberger}}{{Erb} et~al.}{2006}]{Erb06}
{Erb} D.~K.,  {Steidel} C.~C.,  {Shapley} A.~E.,  {Pettini} M.,  {Reddy} N.~A.,
     {Adelberger} K.~L.,  2006, \apj, 646, 107

\bibitem[\protect\citeauthoryear{{Faber} \& {Jackson}}{{Faber} \&
  {Jackson}}{1976}]{FJ}
{Faber} S.~M.,  {Jackson} R.~E.,  1976, \apj, 204, 668

\bibitem[\protect\citeauthoryear{{Fan}, {Lapi}, {De Zotti} \& {Danese}}{{Fan}
  et~al.}{2008}]{Fan08}
{Fan} L.,  {Lapi} A.,  {De Zotti} G.,    {Danese} L.,  2008, \apjl, 689, L101

\bibitem[\protect\citeauthoryear{{Fan}, {Lapi}, {Bressan}, {Bernardi}, {De
  Zotti} \& {Danese}}{{Fan} et~al.}{2010}]{Fan10}
{Fan} L.,  {Lapi} A.,  {Bressan} A.,  {Bernardi} M.,  {De Zotti} G.,
  {Danese} L.,  2010, \apj, 718, 1460

\bibitem[\protect\citeauthoryear{{Fasano}, {Bettoni}, {Ascaso}, {Tormen},
  {Poggianti}, {Valentinuzzi}, {D'Onofrio}, {Fritz}, {Moretti}, {Omizzolo},
  {Cava}, {Moles}, {Dressler}, {Couch}, {Kj{\ae}rgaard} \& {Varela}}{{Fasano}
  et~al.}{2010}]{Fasano10}
{Fasano} G. et al., 2010, \mnras, 404, 1490

\bibitem[\protect\citeauthoryear{{Ferrarese}}{{Ferrarese}}{2002}]{Ferrarese02}
{Ferrarese} L.,  2002, \apj, 578, 90

\bibitem[\protect\citeauthoryear{{Ferrarese} \& {Ford}}{{Ferrarese} \&
  {Ford}}{2005}]{FerrareseFord}
{Ferrarese} L.,  {Ford} H.,  2005, Space Science Reviews, 116, 523

\bibitem[\protect\citeauthoryear{{Fisher} \& {Drory}}{{Fisher} \&
  {Drory}}{2008}]{Fisher08}
{Fisher} D.~B.,  {Drory} N.,  2008, \aj, 136, 773

\bibitem[\protect\citeauthoryear{{Fisher} \& {Drory}}{{Fisher} \&
  {Drory}}{2010}]{Fisher10}
{Fisher} D.~B.,  {Drory} N.,  2010, \apj, 716, 942

\bibitem[\protect\citeauthoryear{{Fisher} \& {Drory}}{{Fisher} \&
  {Drory}}{2011}]{Fisher11}
{Fisher} D.~B.,  {Drory} N.,  2011, \apjl, 733, L47 

\bibitem[\protect\citeauthoryear{{Forbes}, {Spitler}, {Strader}, {Romanowsky},
  {Brodie} \& {Foster}}{{Forbes} et~al.}{2011}]{Forbes11}
{Forbes} D.~A.,  {Spitler} L.~R.,  {Strader} J.,  {Romanowsky} A.~J.,  {Brodie}
  J.~P.,    {Foster} C.,  2011, \mnras, 413, 2943 

\bibitem[\protect\citeauthoryear{{F{\"o}rster Schreiber}, {Shapley}, {Genzel},
  {Bouch{\'e}}, {Cresci}, {Davies}, {Erb}, {Genel}, {Lutz}, {Newman},
  {Shapiro}, {Steidel}, {Sternberg} \& {Tacconi}}{{F{\"o}rster Schreiber}
  et~al.}{2011}]{Forster11}
{F{\"o}rster Schreiber} N.~M. et al., 2011, \apj, 739, 45 


\bibitem[\protect\citeauthoryear{{Foster}, {Proctor}, {Forbes}, {Spolaor},
  {Hopkins} \& {Brodie}}{{Foster} et~al.}{2009}]{Foster09}
{Foster} C.,  {Proctor} R.~N.,  {Forbes} D.~A.,  {Spolaor} M.,  {Hopkins}
  P.~F.,    {Brodie} J.~P.,  2009, \mnras, 400, 2135

\bibitem[\protect\citeauthoryear{{Franx}, {van Dokkum}, {Schreiber}, {Wuyts},
  {Labb{\'e}} \& {Toft}}{{Franx} et~al.}{2008}]{Franx08}
{Franx} M.,  {van Dokkum} P.~G.,  {Schreiber} N.~M.~F.,  {Wuyts} S.,
  {Labb{\'e}} I.,    {Toft} S.,  2008, \apj, 688, 770

\bibitem[\protect\citeauthoryear{{Fu}, {Guo}, {Kauffmann} \& {Krumholz}}{{Fu}
  et~al.}{2010}]{Fu10}
{Fu} J.,  {Guo} Q.,  {Kauffmann} G.,    {Krumholz} M.~R.,  2010, \mnras, 409,
  515

\bibitem[\protect\citeauthoryear{{Gadotti}}{{Gadotti}}{2009}]{Gadotti09}
{Gadotti} D.~A.,  2009, \mnras, 393, 1531

\bibitem[\protect\citeauthoryear{{Gaskell}}{{Gaskell}}{2009}]{GaskellMbhSigma}
{Gaskell} C.~M.,  2009, ArXiv:0908.0328

\bibitem[Genzel et al.(2011)]{Genzel10} Genzel R., Newman S., 
Jones T. et al., 2011, \apj, 733, 101 

\bibitem[\protect\citeauthoryear{{Gonz{\'a}lez}, {Lacey}, {Baugh}, {Frenk} \&
  {Benson}}{{Gonz{\'a}lez} et~al.}{2009}]{Gonzalez09}
{Gonz{\'a}lez} J.~E.,  {Lacey} C.~G.,  {Baugh} C.~M.,  {Frenk} C.~S.,
  {Benson} A.~J.,  2009, \mnras, 397, 1254

\bibitem[\protect\citeauthoryear{{Gonz{\'a}lez}, {Lacey}, {Baugh} \&
  {Frenk}}{{Gonz{\'a}lez} et~al.}{2011}]{Gonzalez11}
{Gonz{\'a}lez} J.~E.,  {Lacey} C.~G.,  {Baugh} C.~M.,    {Frenk} C.~S.,  2011,
  \mnras, 413, 749

\bibitem[\protect\citeauthoryear{{Governato}, {Willman}, {Mayer}, {Brooks},
  {Stinson}, {Valenzuela}, {Wadsley} \& {Quinn}}{{Governato}
  et~al.}{2007}]{Governato07}
{Governato} F.,  {Willman} B.,  {Mayer} L.,  {Brooks} A.,  {Stinson} G.,
  {Valenzuela} O.,  {Wadsley} J.,    {Quinn} T.,  2007, \mnras, 374, 1479

\bibitem[\protect\citeauthoryear{{Granato}, {De Zotti}, {Silva}, {Bressan} \&
  {Danese}}{{Granato} et~al.}{2004}]{Granato04}
{Granato} G.~L.,  {De Zotti} G.,  {Silva} L.,  {Bressan} A.,    {Danese} L.,
  2004, \apj, 600, 580

\bibitem[\protect\citeauthoryear{{Granato}, {Silva}, {Lapi}, {Shankar}, {De
  Zotti} \& {Danese}}{{Granato} et~al.}{2006}]{Granato06}
{Granato} G.~L.,  {Silva} L.,  {Lapi} A.,  {Shankar} F.,  {De Zotti} G.,
  {Danese} L.,  2006, \mnras, 368, L72

\bibitem[\protect\citeauthoryear{{Graves} \& {Faber}}{{Graves} \&
  {Faber}}{2010}]{Graves10}
{Graves} G.~J.,  {Faber} S.~M.,  2010, \apj, 717, 803

\bibitem[Guo et al.(2011)]{Guo11} Guo Q., White S., 
Boylan-Kolchin M. et al., 2011, \mnras, 413, 101 

\bibitem[\protect\citeauthoryear{{H{\"a}ring} \& {Rix}}{{H{\"a}ring} \&
  {Rix}}{2004}]{HaringRix}
{H{\"a}ring} N.,  {Rix} H.-W.,  2004, \apjl, 604, L89

\bibitem[Henriques et al.(2012)]{Henriquez12} Henriques B.~M.~B. et al., 2012, \mnras, 421, 2904 

\bibitem[\protect\citeauthoryear{{Hernquist}}{{Hernquist}}{1990}]{Hernquist90}
{Hernquist} L.,  1990, \apj, 356, 359

\bibitem[\protect\citeauthoryear{{Hernquist}}{{Hernquist}}{1992}]{Hernquist92}
{Hernquist} L.,  1992, \apj, 400, 460

\bibitem[\protect\citeauthoryear{{Ho}}{{Ho}}{2007}]{Ho07}
{Ho} L.~C.,  2007, \apj, 668, 94

\bibitem[\protect\citeauthoryear{{Hopkins}, {Hernquist}, {Cox}, {Keres} \&
  {Wuyts}}{{Hopkins} et~al.}{2009}]{Hop08FP}
{Hopkins} P.~F.,  {Hernquist} L.,  {Cox} T.~J.,  {Keres} D.,    {Wuyts} S.,
  2009, \apj, 691, 1424

\bibitem[\protect\citeauthoryear{{Hopkins}, {Croton}, {Bundy}, {Khochfar}, {van
  den Bosch}, {Somerville}, {Wetzel}, {Keres}, {Hernquist}, {Stewart},
  {Younger}, {Genel} \& {Ma}}{{Hopkins} et~al.}{2010}]{HopkinsMergers}
{Hopkins} P.~F. et al., 2010, \apj, 724,
  915

\bibitem[\protect\citeauthoryear{{Hyde} \& {Bernardi}}{{Hyde} \&
  {Bernardi}}{2009}]{Hyde09a}
{Hyde} J.~B.,  {Bernardi} M.,  2009, \mnras, 394, 1978

\bibitem[Huertas-Company et al.(2012)]{Huertas12} 
Huertas-Company M. et al., 2012, arXiv:1207.5793 

\bibitem[\protect\citeauthoryear{{Kannappan}}{{Kannappan}}{2004}]{Kanna04}
{Kannappan} S.~J.,  2004, \apjl, 611, L89

\bibitem[Kauffmann et al.(2004)]{Kauff04} Kauffmann G. et al., 2004, \mnras, 353, 713 

\bibitem[Kauffmann et al.(2012)]{Kauff12} Kauffmann G. et al., 2012, \mnras, 422, 997 

\bibitem[Khochfar 
\& Burkert(2006)]{KhochfarBurkert} Khochfar S., Burkert A., 2006, \aap, 445, 403 

\bibitem[Khochfar 
\& Silk(2006a)]{Khochfar06a} Khochfar S., Silk J., 2006a, \mnras, 370, 902

\bibitem[\protect\citeauthoryear{{Khochfar} \& {Silk}}{{Khochfar} \&
  {Silk}}{2006b}]{KochfarSilk06Rez}
{Khochfar} S.,  {Silk} J.,  2006b, \apjl, 648, L21

\bibitem[Khochfar 
\& Silk(2009)]{Khochfar09a} Khochfar S., Silk J., 2009, \mnras, 397, 506 

\bibitem[\protect\citeauthoryear{{Khochfar} \& {Silk}}{{Khochfar} \&
  {Silk}}{2011}]{KhochfarSilk11}
{Khochfar} S.,  {Silk} J.,  2011, \mnras, 410, L42

\bibitem[\protect\citeauthoryear{{Kormendy} \& {Kennicutt} Jr.}{{Kormendy} \&
  {Kennicutt}}{2004}]{Kormendy04}
{Kormendy} J.,  {Kennicutt} Jr. R.~C.,  2004, \araa, 42, 603

\bibitem[\protect\citeauthoryear{{La Barbera}, {Busarello}, {Merluzzi}, {de la
  Rosa}, {Coppola} \& {Haines}}{{La Barbera} et~al.}{2008}]{Labarbera08}
{La Barbera} F.,  {Busarello} G.,  {Merluzzi} P.,  {de la Rosa} I.~G.,
  {Coppola} G.,    {Haines} C.~P.,  2008, \apj, 689, 913

\bibitem[\protect\citeauthoryear{{La Barbera} \& {de Carvalho}}{{La Barbera} \&
  {de Carvalho}}{2009}]{LaBarbera09}
{La Barbera} F.,  {de Carvalho} R.~R.,  2009, \apjl, 699, L76

\bibitem[\protect\citeauthoryear{{La Barbera}, {Lopes}, {de Carvalho}, {de La
  Rosa} \& {Berlind}}{{La Barbera} et~al.}{2010}]{LaBarbera10}
{La Barbera} F.,  {Lopes} P.~A.~A.,  {de Carvalho} R.~R.,  {de La Rosa} I.~G.,
    {Berlind} A.~A.,  2010, \mnras, 408, 1361

\bibitem[\protect\citeauthoryear{{Lapi} \& {Cavaliere}}{{Lapi} \&
  {Cavaliere}}{2009}]{Lapi09a}
{Lapi} A.,  {Cavaliere} A.,  2009, \apj, 692, 174

\bibitem[\protect\citeauthoryear{{Loeb} \& {Peebles}}{{Loeb} \&
  {Peebles}}{2003}]{Loeb03}
{Loeb} A.,  {Peebles} P.~J.~E.,  2003, \apj, 589, 29

\bibitem[Magdis et al.(2011)]{Magdis11} Magdis G.~E., Elbaz 
D., Hwang H.~S., Pep Team, \& Hermes Team 2011, Galaxy Evolution: Infrared to Millimeter Wavelength Perspective, 446, 221

\bibitem[\protect\citeauthoryear{{Mancini}, {Matute}, {Cimatti}, {Daddi},
  {Dickinson}, {Rodighiero}, {Bolzonella} \& {Pozzetti}}{{Mancini}
  et~al.}{2009}]{Mancini09}
{Mancini} C.,  {Matute} I.,  {Cimatti} A.,  {Daddi} E.,  {Dickinson} M.,
  {Rodighiero} G.,  {Bolzonella} M.,    {Pozzetti} L.,  2009, \aap, 500, 705

\bibitem[\protect\citeauthoryear{{Mancini}, {Daddi}, {Renzini}, {Salmi},
  {McCracken}, {Cimatti}, {Onodera}, {Salvato}, {Koekemoer}, {Aussel}, {Le
  Floc'h}, {Willott} \& {Capak}}{{Mancini} et~al.}{2010}]{Mancini10}
{Mancini} C. et al., 2010, \mnras, 401, 933

\bibitem[\protect\citeauthoryear{{Marchesini}, {van Dokkum}, {F{\"o}rster
  Schreiber}, {Franx}, {Labb{\'e}} \& {Wuyts}}{{Marchesini}
  et~al.}{2009}]{Marchesini09}
{Marchesini} D.,  {van Dokkum} P.~G.,  {F{\"o}rster Schreiber} N.~M.,  {Franx}
  M.,  {Labb{\'e}} I.,    {Wuyts} S.,  2009, \apj, 701, 1765

\bibitem[\protect\citeauthoryear{{Marchesini}, {Whitaker}, {Brammer}, {van
  Dokkum}, {Labb{\'e}}, {Muzzin}, {Quadri}, {Kriek}, {Lee}, {Rudnick}, {Franx},
  {Illingworth} \& {Wake}}{{Marchesini} et~al.}{2010}]{Marchesini10}
{Marchesini} D. et al., 2010, \apj,
  725, 1277

\bibitem[\protect\citeauthoryear{{Marconi} \& {Hunt}}{{Marconi} \&
  {Hunt}}{2003}]{Marconi03}
{Marconi} A.,  {Hunt} L.~K.,  2003, \apjl, 589, L21

\bibitem[\protect\citeauthoryear{{Marulli}, {Bonoli}, {Branchini}, {Moscardini}
  \& {Springel}}{{Marulli} et~al.}{2008}]{Marulli08}
{Marulli} F.,  {Bonoli} S.,  {Branchini} E.,  {Moscardini} L.,    {Springel}
  V.,  2008, \mnras, 385, 1846

\bibitem[\protect\citeauthoryear{{Marulli}, {Bonoli}, {Branchini}, {Gilli},
  {Moscardini} \& {Springel}}{{Marulli} et~al.}{2009}]{Marulli09}
{Marulli} F.,  {Bonoli} S.,  {Branchini} E.,  {Gilli} R.,  {Moscardini} L.,
  {Springel} V.,  2009, \mnras, 396, 1404

\bibitem[\protect\citeauthoryear{{Menci}, {Cavaliere}, {Fontana}, {Giallongo},
  {Poli} \& {Vittorini}}{{Menci} et~al.}{2004}]{Menci04}
{Menci} N.,  {Cavaliere} A.,  {Fontana} A.,  {Giallongo} E.,  {Poli} F.,
  {Vittorini} V.,  2004, \apj, 604, 12

\bibitem[\protect\citeauthoryear{{Menci}, {Fontana}, {Giallongo}, {Grazian} \&
  {Salimbeni}}{{Menci} et~al.}{2006}]{Menci06}
{Menci} N.,  {Fontana} A.,  {Giallongo} E.,  {Grazian} A.,    {Salimbeni} S.,
  2006, \apj, 647, 753

\bibitem[Merlin et al.(2012)]{Merlin12} Merlin E. et al., 2012, arXiv:1204.5118

\bibitem[\protect\citeauthoryear{{Monaco}, {Fontanot} \& {Taffoni}}{{Monaco}
  et~al.}{2007}]{Monaco07}
{Monaco} P.,  {Fontanot} F.,    {Taffoni} G.,  2007, \mnras, 375, 1189

\bibitem[\protect\citeauthoryear{{Mosleh}, {Williams}, {Franx} \&
  {Kriek}}{{Mosleh} et~al.}{2011}]{Mosleh11}
{Mosleh} M.,  {Williams} R.~J.,  {Franx} M.,    {Kriek} M.,  2011, \apj, 727, 5

\bibitem[Moster et al.(2012)]{Moster12} Moster B.~P., Naab T., White S.~D.~M., 2012, arXiv:1205.5807 

\bibitem[\protect\citeauthoryear{{Naab}, {Jesseit} \& {Burkert}}{{Naab}
  et~al.}{2006}]{Naab06}
{Naab} T.,  {Jesseit} R.,    {Burkert} A.,  2006, \mnras, 372, 839

\bibitem[\protect\citeauthoryear{{Naab}, {Johansson}, {Ostriker} \&
  {Efstathiou}}{{Naab} et~al.}{2007}]{Naab07paper}
{Naab} T.,  {Johansson} P.~H.,  {Ostriker} J.~P.,    {Efstathiou} G.,  2007,
  \apj, 658, 710

\bibitem[\protect\citeauthoryear{{Naab}, {Johansson} \& {Ostriker}}{{Naab}
  et~al.}{2009}]{Naab09}
{Naab} T.,  {Johansson} P.~H.,    {Ostriker} J.~P.,  2009, \apjl, 699, L178

\bibitem[\protect\citeauthoryear{{Naab} \& {Ostriker}}{{Naab} \&
  {Ostriker}}{2009}]{Naab07}
{Naab} T.,  {Ostriker} J.~P.,  2009, \apj, 690, 1452 

\bibitem[\protect\citeauthoryear{{Nair}, {van den Bergh} \& {Abraham}}{{Nair}
  et~al.}{2010}]{Nair10}
{Nair} P.~B.,  {van den Bergh} S.,    {Abraham} R.~G.,  2010, \apj, 715, 606

\bibitem[\protect\citeauthoryear{{Nair}, {van den Bergh} \& {Abraham}}{{Nair}
  et~al.}{2011}]{Nair11}
{Nair} P.,  {van den Bergh} S.,    {Abraham} R.~G.,  2011, \apjl, 734, L31

\bibitem[\protect\citeauthoryear{{Napolitano}, {Romanowsky} \&
  {Tortora}}{{Napolitano} et~al.}{2010}]{Napolitano10}
{Napolitano} N.~R.,  {Romanowsky} A.~J.,    {Tortora} C.,  2010, \mnras, 405,
  2351

\bibitem[\protect\citeauthoryear{{Navarro}, {Frenk} \& {White}}{{Navarro}
  et~al.}{1997}]{NFW}
{Navarro} J.~F.,  {Frenk} C.~S.,    {White} S.~D.~M.,  1997, \apj, 490, 493

\bibitem[Neistein 
\& Weinmann(2010)]{NeisteinWein} Neistein E., Weinmann S.~M., 2010, \mnras, 405, 2717 

\bibitem[\protect\citeauthoryear{{Neistein}, {Li}, {Khochfar}, {Weinmann},
  {Shankar} \& {Boylan-Kolchin}}{{Neistein} et~al.}{2011}]{Neistein11}
{Neistein} E.,  {Li} C.,  {Khochfar} S.,  {Weinmann} S.~M.,  {Shankar} F.,
  {Boylan-Kolchin} M.,  2011, \mnras, 416, 1486

\bibitem[\protect\citeauthoryear{{Newman}, {Ellis}, {Bundy} \& {Treu}}{{Newman}
  et~al.}{2012}]{Newman12}
{Newman} A.~B.,  {Ellis} R.~S.,  {Bundy} K.,    {Treu} T.,  2012, \apj, 746,
  162

\bibitem[Newton et al.(2011)]{Newton11} Newton E.~R., Marshall
P.~J., Treu T. et al., 2011, \apj, 734, 104 

\bibitem[\protect\citeauthoryear{{Nipoti}, {Treu} \& {Bolton}}{{Nipoti}
  et~al.}{2009}]{Nipoti09}
{Nipoti} C.,  {Treu} T.,    {Bolton} A.~S.,  2009, \apj, 706, 86

\bibitem[Nipoti et al.(2012)]{Nipoti12} Nipoti C., Treu T., 
Leauthaud A. et al., 2012, \mnras, 422, 1714 

\bibitem[\protect\citeauthoryear{{Oser}, {Ostriker}, {Naab}, {Johansson} \&
  {Burkert}}{{Oser} et~al.}{2010}]{Oser10}
{Oser} L.,  {Ostriker} J.~P.,  {Naab} T.,  {Johansson} P.~H.,    {Burkert} A.,
  2010, \apj, 725, 2312

\bibitem[Papovich et al.(2012)]{Papovich11} Papovich C., Bassett
R., Lotz J.~M. et al., 2012, \apj, 750, 93 

\bibitem[\protect\citeauthoryear{{Peeples} \& {Shankar}}{{Peeples} \&
  {Shankar}}{2010}]{PeeplesShankar}
{Peeples} M.~S.,  {Shankar} F.,  2011, \mnras, 417, 2962 

\bibitem[\protect\citeauthoryear{{Pizzella}, {Corsini}, {Dalla Bont{\`a}},
  {Sarzi}, {Coccato} \& {Bertola}}{{Pizzella} et~al.}{2005}]{Pizzella05}
{Pizzella} A.,  {Corsini} E.~M.,  {Dalla Bont{\`a}} E.,  {Sarzi} M.,  {Coccato}
  L.,    {Bertola} F.,  2005, \apj, 631, 785

\bibitem[\protect\citeauthoryear{{Prugniel} \& {Simien}}{{Prugniel} \&
  {Simien}}{1997}]{Prugniel97}
{Prugniel} P.,  {Simien} F.,  1997, \aap, 321, 111

\bibitem[\protect\citeauthoryear{{Ragone-Figueroa} \&
  {Granato}}{{Ragone-Figueroa} \& {Granato}}{2011}]{RagoneGranato}
{Ragone-Figueroa} C.,  {Granato} G.~L.,  2011, \mnras, 414, 3690

\bibitem[\protect\citeauthoryear{{Raichoor}, {Mei}, {Stanford}, {Holden},
  {Nakata}, {Rosati}, {Shankar}, {Tanaka}, {Ford}, {Huertas-Company},
  {Illingworth}, {Kodama}, {Postman}, {Rettura}, {Blakeslee}, {Demarco}, {Jee}
  \& {White}}{{Raichoor} et~al.}{2012}]{Raichoor12}
{Raichoor} A.,  et al.,  2012, \apj, 745, 130

\bibitem[\protect\citeauthoryear{{Robertson}, {Bullock}, {Cox}, {Di Matteo},
  {Hernquist}, {Springel} \& {Yoshida}}{{Robertson} et~al.}{2006}]{Robertson06}
{Robertson} B.,  {Bullock} J.~S.,  {Cox} T.~J.,  {Di Matteo} T.,  {Hernquist}
  L.,  {Springel} V.,    {Yoshida} N.,  2006, \apj, 645, 986

\bibitem[Ryan et al.(2012)]{Ryan10} Ryan R.~E. Jr., 
McCarthy P.~J., Cohen S.~H., et al., 2012, \apj, 749, 53 

\bibitem[\protect\citeauthoryear{{Saracco}, {Longhetti} \& {Andreon}}{{Saracco}
  et~al.}{2008}]{Saracco08}
{Saracco} P.,  {Longhetti} M.,    {Andreon} S.,  2009, \mnras, 392, 718

\bibitem[\protect\citeauthoryear{{Saracco}, {Longhetti} \&
  {Gargiulo}}{{Saracco} et~al.}{2010}]{Saracco10}
{Saracco} P.,  {Longhetti} M.,    {Gargiulo} A.,  2010, \mnras, 408, L21

\bibitem[\protect\citeauthoryear{{Saracco}, {Longhetti} \&
  {Gargiulo}}{{Saracco} et~al.}{2011}]{Saracco11}
{Saracco} P.,  {Longhetti} M.,    {Gargiulo} A.,  2011, \mnras, 412, 2707

\bibitem[\protect\citeauthoryear{{S{\'e}rsic}}{{S{\'e}rsic}}{1963}]{Sersic63}
{S{\'e}rsic} J.~L.,  1963, Boletin de la Asociacion Argentina de Astronomia La
  Plata Argentina, 6, 41

\bibitem[\protect\citeauthoryear{{Shankar}, {Salucci}, {Granato}, {De Zotti} \&
  {Danese}}{{Shankar} et~al.}{2004}]{Shankar04}
{Shankar} F.,  {Salucci} P.,  {Granato} G.~L.,  {De Zotti} G.,    {Danese} L.,
  2004, \mnras, 354, 1020

\bibitem[\protect\citeauthoryear{{Shankar}, {Lapi}, {Salucci}, {De Zotti} \&
  {Danese}}{{Shankar} et~al.}{2006}]{Shankar06}
{Shankar} F.,  {Lapi} A.,  {Salucci} P.,  {De Zotti} G.,    {Danese} L.,  2006,
  \apj, 643, 14

\bibitem[\protect\citeauthoryear{{Shankar}}{{Shankar}}{2009}]{ShankarReview}
{Shankar} F.,  2009, New Astr. Rev., 53, 57 

\bibitem[\protect\citeauthoryear{{Shankar} \& {Bernardi}}{{Shankar} \&
  {Bernardi}}{2009}]{ShankarBernardi}
{Shankar} F.,  {Bernardi} M.,  2009, \mnras, 396, L76

\bibitem[\protect\citeauthoryear{{Shankar}, {Bernardi} \& {Haiman}}{{Shankar}
  et~al.}{2009}]{ShankarMsigma}
{Shankar} F.,  {Bernardi} M.,    {Haiman} Z.,  2009, \apj, 694, 867

\bibitem[\protect\citeauthoryear{{Shankar}, {Marulli}, {Bernardi},
  {Boylan-Kolchin}, {Dai} \& {Khochfar}}{{Shankar} et~al.}{2010a}]{ShankarPhire}
{Shankar} F.,  {Marulli} F.,  {Bernardi} M.,  {Boylan-Kolchin} M.,  {Dai} X.,
   {Khochfar} S.,  2010a, \mnras, 405, 948

\bibitem[\protect\citeauthoryear{{Shankar}, {Marulli}, {Bernardi}, {Dai},
  {Hyde} \& {Sheth}}{{Shankar} et~al.}{2010b}]{ShankarRe}
{Shankar} F.,  {Marulli} F.,  {Bernardi} M.,  {Dai} X.,  {Hyde} J.~B.,
  {Sheth} R.~K.,  2010b, \mnras, 403, 117


\bibitem[\protect\citeauthoryear{{Shankar}, {Marulli}, {Mathur}, {Bernardi} \&
  {Bournaud}}{{Shankar} et~al.}{2012}]{Shankar12}
{Shankar} F.,  {Marulli} F.,  {Mathur} S.,  {Bernardi} M.,    {Bournaud} F.,
  2012, \aap, 540, A23

\bibitem[\protect\citeauthoryear{{Shen}, {Mo}, {White}, {Blanton}, {Kauffmann},
  {Voges}, {Brinkmann} \& {Csabai}}{{Shen} et~al.}{2003}]{Shen03}
{Shen} S.,  {Mo} H.~J.,  {White} S.~D.~M.,  {Blanton} M.~R.,  {Kauffmann} G.,
  {Voges} W.,  {Brinkmann} J.,    {Csabai} I.,  2003, \mnras, 343, 978

\bibitem[\protect\citeauthoryear{{Sheth}, {Bernardi}, {Schechter}, {Burles},
  {Eisenstein}, {Finkbeiner}, {Frieman}, {Lupton}, {Schlegel}, {Subbarao},
  {Shimasaku}, {Bahcall}, {Brinkmann} \& {Ivezi{\'c}}}{{Sheth}
  et~al.}{2003}]{sheth03}
{Sheth} R.~K., et al. 2003, \apj, 594, 225

\bibitem[\protect\citeauthoryear{{Shields}, {Menezes}, {Massart} \& {Vanden
  Bout}}{{Shields} et~al.}{2006}]{Shields06}
{Shields} G.~A.,  {Menezes} K.~L.,  {Massart} C.~A.,    {Vanden Bout} P.,
  2006, \apj, 641, 683

\bibitem[\protect\citeauthoryear{{Somerville}, {Barden}, {Rix}, {Bell},
  {Beckwith}, {Borch}, {Caldwell}, {H{\"a}u{\ss}ler}, {Heymans}, {Jahnke},
  {Jogee}, {McIntosh}, {Meisenheimer}, {Peng}, {S{\'a}nchez}, {Wisotzki} \&
  {Wolf}}{{Somerville} et~al.}{2008}]{Somerville08}
{Somerville} R.~S. et al., 2008, \apj, 672, 776

\bibitem[\protect\citeauthoryear{{Somerville}, {Hopkins}, {Cox}, {Robertson} \&
  {Hernquist}}{{Somerville} et~al.}{2008}]{Somerville08SAM}
{Somerville} R.~S.,  {Hopkins} P.~F.,  {Cox} T.~J.,  {Robertson} B.~E.,
  {Hernquist} L.,  2008, \mnras, 391, 481

\bibitem[\protect\citeauthoryear{{Spolaor}, {Hau}, {Forbes} \&
  {Couch}}{{Spolaor} et~al.}{2010}]{Spolaor10b}
{Spolaor} M.,  {Hau} G.~K.~T.,  {Forbes} D.~A.,    {Couch} W.~J.,  2010,
  \mnras, 408, 254

\bibitem[\protect\citeauthoryear{{Springel}, {White}, {Jenkins}, {Frenk},
  {Yoshida}, {Gao}, {Navarro}, {Thacker}, {Croton}, {Helly}, {Peacock}, {Cole},
  {Thomas}, {Couchman}, {Evrard}, {Colberg} \& {Pearce}}{{Springel}
  et~al.}{2005}]{Springel05b}
{Springel} V. et al.,  2005, \nat, 435, 629

\bibitem[\protect\citeauthoryear{{Stewart}, {Bullock}, {Wechsler}, {Maller} \&
  {Zentner}}{{Stewart} et~al.}{2008}]{Stewart08}
{Stewart} K.~R.,  {Bullock} J.~S.,  {Wechsler} R.~H.,  {Maller} A.~H.,
  {Zentner} A.~R.,  2008, \apj, 683, 597

\bibitem[\protect\citeauthoryear{{Stott}, {Collins}, {Burke}, {Hamilton-Morris}
  \& {Smith}}{{Stott} et~al.}{2011}]{Stott11}
{Stott} J.~P.,  {Collins} C.~A.,  {Burke} C.,  {Hamilton-Morris} V.,    {Smith}
  G.~P.,  2011, \mnras, 414, 445

\bibitem[\protect\citeauthoryear{{Szomoru}, {Franx}, {van Dokkum}, {Trenti},
  {Illingworth}, {Labb{\'e}}, {Bouwens}, {Oesch} \& {Carollo}}{{Szomoru}
  et~al.}{2010}]{Szomoru10}
{Szomoru} D. et al., 2010, \apjl, 714, L244

\bibitem[\protect\citeauthoryear{{Franx}, {van Dokkum}, {Schreiber}, {Wuyts},
  {Labb{\'e}} \& {Toft}}{{Tacconi} et~al.}{2008}]{Tacconi08}
{Tacconi} L.~J. et al., 2008, \apj, 680, 246

\bibitem[\protect\citeauthoryear{{Taylor}, {Franx}, {Glazebrook}, {Brinchmann},
  {van der Wel} \& {van Dokkum}}{{Taylor} et~al.}{2010}]{Taylor10}
{Taylor} E.~N.,  {Franx} M.,  {Glazebrook} K.,  {Brinchmann} J.,  {van der Wel}
  A.,    {van Dokkum} P.~G.,  2010, \apj, 720, 723

\bibitem[\protect\citeauthoryear{{Terzi{\'c}} \& {Graham}}{{Terzi{\'c}} \&
  {Graham}}{2005}]{TerzicGraham}
{Terzi{\'c}} B.,  {Graham} A.~W.,  2005, \mnras, 362, 197

\bibitem[\protect\citeauthoryear{{Tissera}, {White}, {Pedrosa} \&
  {Scannapieco}}{{Tissera} et~al.}{2010}]{Tissera10}
{Tissera} P.~B.,  {White} S.~D.~M.,  {Pedrosa} S.,    {Scannapieco} C.,  2010,
  \mnras, 406, 922

\bibitem[\protect\citeauthoryear{{Tortora}, {Napolitano}, {Romanowsky},
  {Capaccioli} \& {Covone}}{{Tortora} et~al.}{2009}]{Tortora09}
{Tortora} C.,  {Napolitano} N.~R.,  {Romanowsky} A.~J.,  {Capaccioli} M.,
  {Covone} G.,  2009, \mnras, 396, 1132

\bibitem[Tortora et al.(2012)]{Tortora12} Tortora C., 
Romanowsky A.~J., Napolitano N.~R., 2012, arXiv:1207.4475 

\bibitem[\protect\citeauthoryear{{Treu}, {Woo}, {Malkan} \& {Blandford}}{{Treu}
  et~al.}{2007}]{Treu07}
{Treu} T.,  {Woo} J.-H.,  {Malkan} M.~A.,    {Blandford} R.~D.,  2007, \apj,
  667, 117

\bibitem[\protect\citeauthoryear{{Trujillo}, {Feulner}, {Goranova}, {Hopp},
  {Longhetti}, {Saracco}, {Bender}, {Braito}, {Della Ceca}, {Drory}, {Mannucci}
  \& {Severgnini}}{{Trujillo} et~al.}{2006}]{Trujillo06}
{Trujillo} I., et al. 2006, \mnras, 373, L36

\bibitem[\protect\citeauthoryear{{Trujillo}, {Conselice}, {Bundy}, {Cooper},
  {Eisenhardt} \& {Ellis}}{{Trujillo} et~al.}{2007}]{Trujillo07}
{Trujillo} I.,  {Conselice} C.~J.,  {Bundy} K.,  {Cooper} M.~C.,  {Eisenhardt}
  P.,    {Ellis} R.~S.,  2007, \mnras, 382, 109

\bibitem[\protect\citeauthoryear{{Trujillo}, {Cenarro}, {de
  Lorenzo-C{\'a}ceres}, {Vazdekis}, {de la Rosa} \& {Cava}}{{Trujillo}
  et~al.}{2009}]{Trujillo09}
{Trujillo} I.,  {Cenarro} A.~J.,  {de Lorenzo-C{\'a}ceres} A.,  {Vazdekis} A.,
  {de la Rosa} I.~G.,    {Cava} A.,  2009, \apjl, 692, L118

\bibitem[\protect\citeauthoryear{{Trujillo}, {Ferreras} \& {de la
  Rosa}}{{Trujillo} et~al.}{2011}]{Trujillo11}
{Trujillo} I.,  {Ferreras} I.,    {de la Rosa} I.~G.,  2011, \mnras, 415, 3903

\bibitem[\protect\citeauthoryear{{Tundo}, {Bernardi}, {Hyde}, {Sheth} \&
  {Pizzella}}{{Tundo} et~al.}{2007}]{Tundo07}
{Tundo} E.,  {Bernardi} M.,  {Hyde} J.~B.,  {Sheth} R.~K.,    {Pizzella} A.,
  2007, \apj, 663, 53

\bibitem[\protect\citeauthoryear{{Valentinuzzi}, {Fritz}, {Poggianti}, {Cava},
  {Bettoni}, {Fasano}, {D'Onofrio}, {Couch}, {Dressler}, {Moles}, {Moretti},
  {Omizzolo}, {Kj{\ae}rgaard}, {Vanzella} \& {Varela}}{{Valentinuzzi}
  et~al.}{2010a}]{Valentinuzzi10a}
{Valentinuzzi} T.  et al., 2010, \apj, 712, 226

\bibitem[\protect\citeauthoryear{{Valentinuzzi}, {Poggianti}, {Saglia},
  {Arag{\'o}n-Salamanca}, {Simard}, {S{\'a}nchez-Bl{\'a}zquez}, {D'onofrio},
  {Cava}, {Couch}, {Fritz}, {Moretti} \& {Vulcani}}{{Valentinuzzi}
  et~al.}{2010b}]{Valentinuzzi10b}
{Valentinuzzi} T. et al., 2010,
  \apjl, 721, L19

\bibitem[van de Sande et al.(2011)]{Sande11} van de Sande J. et al., 2011, \apjl, 736, L9 

\bibitem[\protect\citeauthoryear{{van der Wel}, {Bell}, {van den Bosch},
  {Gallazzi} \& {Rix}}{{van der Wel} et~al.}{2009}]{vanderwel09}
{van der Wel} A.,  {Bell} E.~F.,  {van den Bosch} F.~C.,  {Gallazzi} A.,
  {Rix} H.,  2009, \apj, 698, 1232

\bibitem[\protect\citeauthoryear{{van der Wel}, {Rix}, {Wuyts}, {McGrath},
  {Koekemoer}, {Bell}, {Holden}, {Robaina} \& {McIntosh}}{{van der Wel}
  et~al.}{2011}]{vanderwel11}
{van der Wel} A. et al.,  2011, \apj, 730, 38

\bibitem[\protect\citeauthoryear{{van der Wel}, {Holden}, {Zirm}, {Franx},
  {Rettura}, {Illingworth} \& {Ford}}{{van der Wel} et~al.}{2008}]{Vanderwel08}
{van der Wel} A.,  {Holden} B.~P.,  {Zirm} A.~W.,  {Franx} M.,  {Rettura} A.,
  {Illingworth} G.~D.,    {Ford} H.~C.,  2008, \apj, 688, 48

\bibitem[\protect\citeauthoryear{{van Dokkum}, {Franx}, {Kriek}, {Holden},
  {Illingworth}, {Magee}, {Bouwens}, {Marchesini}, {Quadri}, {Rudnick},
  {Taylor} \& {Toft}}{{van Dokkum} et~al.}{2008}]{Vandokkum08}
{van Dokkum} P.~G., et al. 2008, \apjl, 677, L5

\bibitem[\protect\citeauthoryear{{van Dokkum}, {Kriek} \& {Franx}}{{van Dokkum}
  et~al.}{2009}]{vandokkum09}
{van Dokkum} P.~G.,  {Kriek} M.,    {Franx} M.,  2009, \nat, 460, 717

\bibitem[\protect\citeauthoryear{{van Dokkum}, {Whitaker}, {Brammer}, {Franx},
  {Kriek}, {Labb{\'e}}, {Marchesini}, {Quadri}, {Bezanson}, {Illingworth},
  {Muzzin}, {Rudnick}, {Tal} \& {Wake}}{{van Dokkum}
  et~al.}{2010}]{VanDokkum10}
{van Dokkum} P.~G. et al., 2010, \apj, 709, 1018

\bibitem[\protect\citeauthoryear{{Wang}, {Navarro}, {Frenk}, {White},
  {Springel}, {Jenkins}, {Helmi}, {Ludlow} \& {Vogelsberger}}{{Wang}
  et~al.}{2011}]{Wang11}
{Wang} J. et al.,  2011, \mnras, 413, 1373

\bibitem[Weinmann et al.(2009)]{Weinmann09} Weinmann S.~M., 
Kauffmann G., van den Bosch F.~C., et al., 2009, \mnras, 394, 1213 

\bibitem[\protect\citeauthoryear{{Weinmann}, {Neistein} \& {Dekel}}{{Weinmann}
  et~al.}{2011}]{Weinmann11}
{Weinmann} S.~M.,  {Neistein} E.,    {Dekel} A.,  2011, \mnras, 417, 2737 

\bibitem[\protect\citeauthoryear{{Whitaker}, {Kriek}, {van Dokkum}, {Bezanson},
  {Brammer}, {Franx} \& {Labb{\'e}}}{{Whitaker} et~al.}{2012}]{Whitaker12}
{Whitaker} K.~E.,  {Kriek} M.,  {van Dokkum} P.~G.,  {Bezanson} R.,  {Brammer}
  G.,  {Franx} M.,    {Labb{\'e}} I.,  2012, \apj, 745, 179

\bibitem[Williams et al.(2010)]{Williams09} Williams R.~J. et al., 2010, \apj, 713, 738

\bibitem[\protect\citeauthoryear{{Woo}, {Treu}, {Malkan} \& {Blandford}}{{Woo}
  et~al.}{2006}]{Woo06}
{Woo} J.-H.,  {Treu} T.,  {Malkan} M.~A.,    {Blandford} R.~D.,  2006, \apj,
  645, 900

\bibitem[\protect\citeauthoryear{{Woo}, {Treu}, {Malkan} \& {Blandford}}{{Woo}
  et~al.}{2008}]{Woo08}
{Woo} J.-H.,  {Treu} T.,  {Malkan} M.~A.,    {Blandford} R.~D.,  2008, \apj,
  681, 925

\bibitem[Yang et al.(2007)]{Yang07} Yang X., Mo H.~J. et al., 2007, \apj, 671, 153 

\bibitem[\protect\citeauthoryear{{Younger}, {Fazio}, {Wilner}, {Ashby},
  {Blundell}, {Gurwell}, {Huang}, {Iono}, {Peck}, {Petitpas}, {Scott}, {Wilson}
  \& {Yun}}{{Younger} et~al.}{2008}]{Younger08}
{Younger} J.~D. et al.,  2008, \apj, 688, 59

\bibitem[\protect\citeauthoryear{{Zhao}, {Mo}, {Jing} \& {B{\"o}rner}}{{Zhao}
  et~al.}{2003}]{Zhao03}
{Zhao} D.~H.,  {Mo} H.~J.,  {Jing} Y.~P.,    {B{\"o}rner} G.,  2003, \mnras,
  339, 12

\end{thebibliography}

\appendix

\section{Relation with Age}
\label{sec|sizeage}

\begin{figure*}
    \includegraphics[width=15truecm]{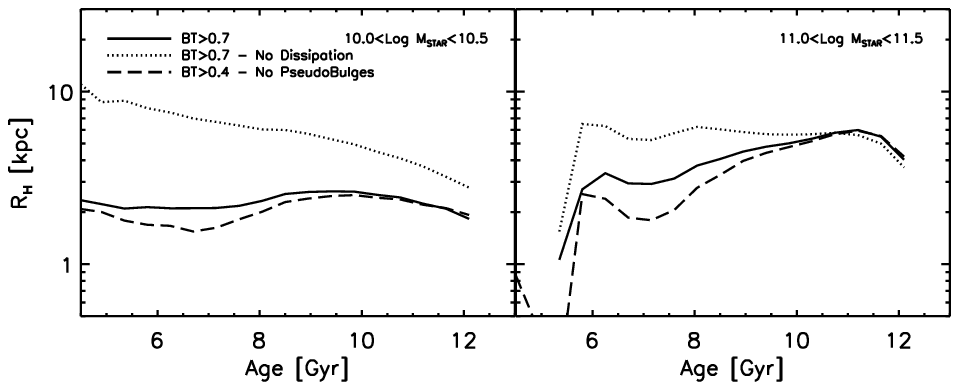}
    \caption{Predicted median size of galaxies of a given stellar mass as a function of mass-weighted age for
    different $B/T$ thresholds, and with and without dissipation included, as labelled. This relation for
    bulge-dominated galaxies is flat at fixed stellar mass, as observed in the local Universe \citep[e.g.,][]{ShankarRe}.}
    \label{fig|ReAge}
\end{figure*}

Although a large number of massive early-type galaxies
at high redshifts have been observed to be more compact
and with higher velocity dispersion than their local counterparts, still
a significant
fraction shows to be already evolved.
\citet{Mancini09} suggested a downsizing scenario
in sizes, with the most massive galaxies approaching the local size-mass
relation earlier than less massive ones.
\citet{Cappellari09}
also discussed that a large fraction of their galaxies at
$1.4 \lesssim z \lesssim 2$ with velocity dispersion from stacked spectrum,
are consistent to the most dense local galaxies of the same mass.
\citet{Saracco11} showed that at the average redshift of $z\sim 1.5$
older galaxies at fixed stellar mass tend
to lie a factor of $\sim 2-3$ below
the size-stellar mass relation characterizing
local early-type galaxies, while younger galaxies
are consistent with it. The latter type of observations
introduced the concept of \emph{assembly bias}, i.e.,
the youngest galaxies at any epoch may be larger than their
older counterparts of similar stellar mass \citep[but see][]{Whitaker12}.

In the local Universe the assembly bias
seems instead to be erased, at least above $\mstare \sim 10^{10}\, \msune$ and large \bt.
\citet{ShankarBernardi}, \citet{ShankarRe},
\citet{Bernardi10} showed from a large sample
of early-type galaxies extracted from SDSS that old and young galaxies of similar stellar
mass share similar sizes, i.e., the size-age relation is rather flat.
\citet{Trujillo11}, more recently, confirmed these results showing that
the size-age relation is flat already at $z\sim 1$.
When moving to lenticular galaxies with lower \bt\ the relation gets tilted,
with the more compact bulges being the oldest \citep[e.g.,][]{vanderwel09,Bernardi10}.
Whatever mechanism puffs up spheroids must be fine-tuned to allow all
galaxies of a given stellar mass to end up on the same size-mass and size-age relations in the local Universe.

Preliminary theoretical studies to interpret the size-age relation
were performed by \citet{ShankarRe} within the context of
the \citet{Bower06} hierarchical model.
Their study showed that because older galaxies usually undergo more mergers,
they naturally grow more and are able to ``catch up'' with the younger ones
producing a rather flat size-age relation.
Here we want to extend the preliminary study by \citet{ShankarRe}
to the case of the present hierarchical model that, at variance
with the Bower et al. one, includes dissipation and better matches
the local size-mass relation.
We will focus here only on the predicted size-age
relation in the local Universe. At redshifts $z\gtrsim 1-1.5$
the spread in ages becomes too small at fixed stellar mass and,
along with its large scatter, does not provide any meaningful prediction to compare with the data.

Figure~\ref{fig|ReAge} shows the predicted median size of galaxies of a given stellar mass as a function of
mass-weighted age for different \bt\ thresholds. For lower masses (left panel) the model predicts that
in the absence of dissipation (dotted lines) more compact galaxies are also the oldest ones.
This is mainly induced by redshift evolution in the sizes of the progenitors
that shrinking at higher redshifts (Figure~\ref{fig|ReMsRedshiftAll}) produce more compact early-type remnants.
Including dissipation (solid lines) flattens the size-age relation out.
As discussed above, dissipation acts in a way to reduce sizes proportionally
to the gas fractions in the progenitors. Within a given mass bin,
the lower mass galaxies have, on average, higher \fgas\ thus suffering more
dissipation than the galaxies lying closer to the more massive end of the bin.
Thus, in this model both mergers and dissipation act in a coordinated way to flatten out the relation.

At higher masses (right panel) the inverse relation of sizes with age disappears,
even in the absence of dissipation. In fact, including dissipation slightly steepens
the relation. These results are all in broad agreement with what observed in SDSS.
However, there are also some discrepancies. As anticipated above,
data on lenticulars, or in general galaxies with
lower $B/T$ ratios, seem to follow a different trend, with older galaxies
being more compact \citep{Bernardi10}. As seen
in Figure~\ref{fig|ReAge}, however, the model seems not to show any significant
change in the shape of the size-age relation when lowering the $B/T$ threshold
(long-dashed lines; only galaxies with bulges mainly grown through mergers are considered here).
This inconsistent behaviour needs to be further understood both observationally and theoretically.

\section{Properties of Pseudobulges}
\label{sec|sizemasspseudo}

There is increasing empirical evidence that
not all bulges in the local Universe can
be formed via mergers \citep[e.g.,][]{Kormendy04,Fisher08,Gadotti09}.
Recent works suggest in fact that a large fraction, possibly the majority of local bulges
are pseudobulges \citep{Fisher11}, with well defined properties different
from those of classical bulges of similar mass. Pseudobulges are usually characterized by younger
stellar populations, they are usually rotation rather than pressure supported,
have less concentrated
surface brightness profiles, and tend to follow well distinct scaling relations
in their global structural properties with respect to classical bulges.

\begin{figure}
    \includegraphics[width=8.7truecm]{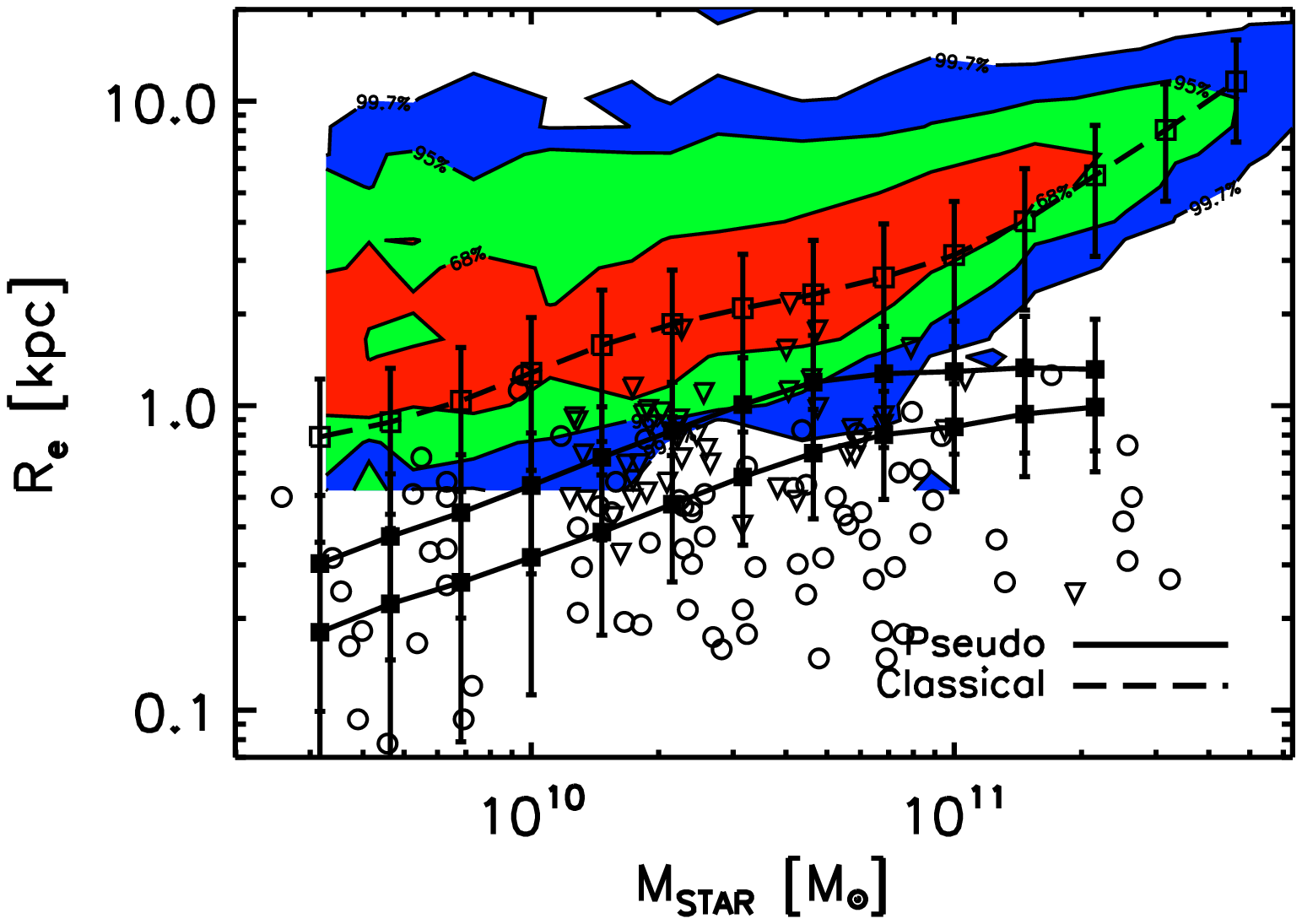}
    \caption{Predicted median effective radius as a function of stellar mass for bulges with sizes
    grown mainly by mergers (\emph{long-dashed} line with their $1-\sigma$ dispersion),
    and bulges mainly grown via disc instability (\emph{solid} lines).
    The classical bulges are shown for \bt$>0.5$, while
    the \emph{upper} and \emph{lower} \emph{solid} lines refer to pseudobulges
    with \bt$>0.2$ and \bt$>0.5$, respectively.
    The \emph{colored contours} are the SDSS subsample with corresponding \bt$>0.5$,
    while the \emph{triangles} and \emph{circles} are pseudobulges measurements from \citet{Gadotti09} and \citet{Fisher10},
    respectively, with no cut on \bt.}
    \label{fig|ReMsPseudo}
\end{figure}

In particular, pseudobulges are found to be much more compact
by a factor of a few at fixed stellar mass with respect to classical bulges.
The open triangles and circles in Figure~\ref{fig|ReMsPseudo} are the measured sizes of
pseudobulges from the samples of \citet{Gadotti09} and \citet{Fisher10}, respectively.
Both their stellar mass measurements have been converted to a common initial
mass function using the tabulated values in \citet{Bernardi10}.
It is interesting to note that despite the completely independent and
different methods utilized, these groups agree in finding that
pseudobulges are more compact by a factor of a few.

The model predicts a size-mass relation for pseudobulges (solid lines and filled squares) -
that we here label as those bulges that have grown more than 50\% of their
final size via disc instability - with a slope and scatter
similar to the one predicted for classical bulges (long-dashed line and open squares)
but lower in normalization by a factor of a few. The upper and lower solid lines referring to pseudobulges
with \bt$>0.2$ and \bt$>0.5$, respectively, are compared with
data by \citet[][triangles]{Gadotti09} and \citet[][circles]{Fisher10},
with no cut on \bt. The predictions are in good agreement with the data, and even more so
given that no fine-tuning was imposed in the model parameters. The model
also predicts that no pseudobulges are found above
$\mstare \gtrsim 2\times 10^{11}\msune$, again in line with the data.
As discussed above, we stress that most of the pseudobulges produced
in the model have \bt$<0.7$.

\begin{figure}
    \includegraphics[width=8.7truecm]{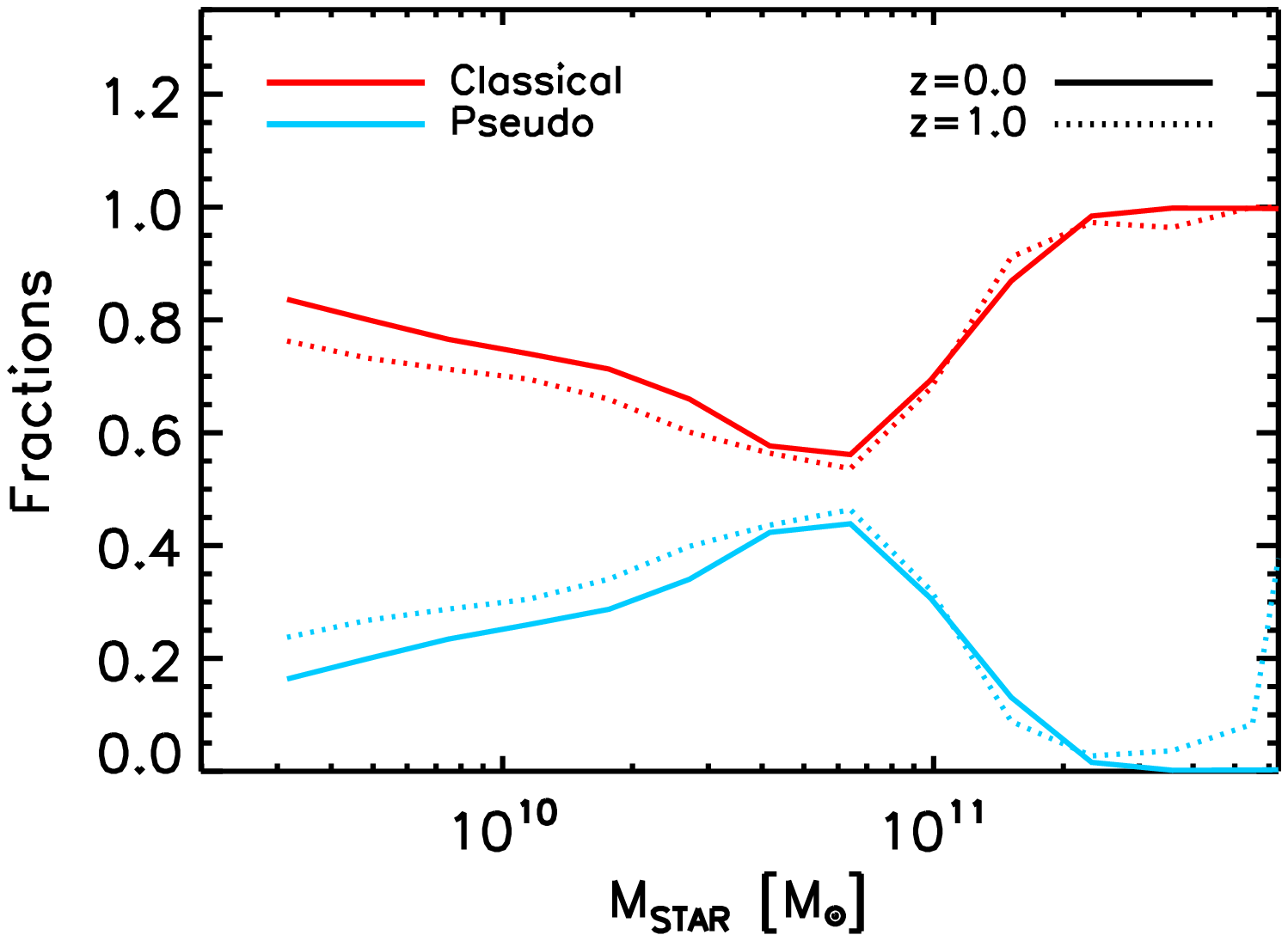}
    \caption{Fraction of pseudo (\emph{cyan} lines) and classical (\emph{red} lines) bulges as a function of stellar mass for redshift $z=0$ (\emph{solid} lines) and $z=1$ (\emph{dotted} lines).
    There is a tendency to have an increasing fraction of pseudobulges at lower stellar masses in broad agreement with the data.}
    \label{fig|FractionPseudo}
\end{figure}

More specifically, Figure\ref{fig|FractionPseudo}
shows that the fraction of all pseudobulges (cyan lines)
and classical (red lines) bulges (here all bulges are considered)
are a non-trivial function of stellar mass.
The fraction of pseudobulges tends to peak around
$\mstare \sim 5\times 10^{10}\, \msune$, and sharply
decreases at higher masses, where the contribution of
classical bulges dominates. This behaviour is in good agreement
with the recent study by \citet[][see also Kormendy et al. 2010]{Fisher11}, who presented
an inventory of galaxy bulge types in a volume-limited
sample within the local 11 Mpc volume using Spitzer and
HST data (see their Figure 3). Theoretically, this is
expected because mergers dominate the size evolution above
the characteristic mass of $10^{11}\, \msune$,
as anticipated in Section~\ref{subsec|firstcomparison}
and further developed in Section~\ref{subsec|evolution}.
On the other hand, \citet{Fisher11} also claim that
pseudobulges tend to be the dominant class of bulges
at lower masses, being close to 60\% around $\mstare \sim 10^{10}\, \msune$.
The model instead predicts a lowering of the fraction of pseudobulges towards lower masses.
More detailed observational and theoretical studies of pseudobulges in the low-mass regime,
beyond the scope of the present paper, are needed to fully understand this discrepancy.

\begin{figure*}
    \includegraphics[width=15.truecm]{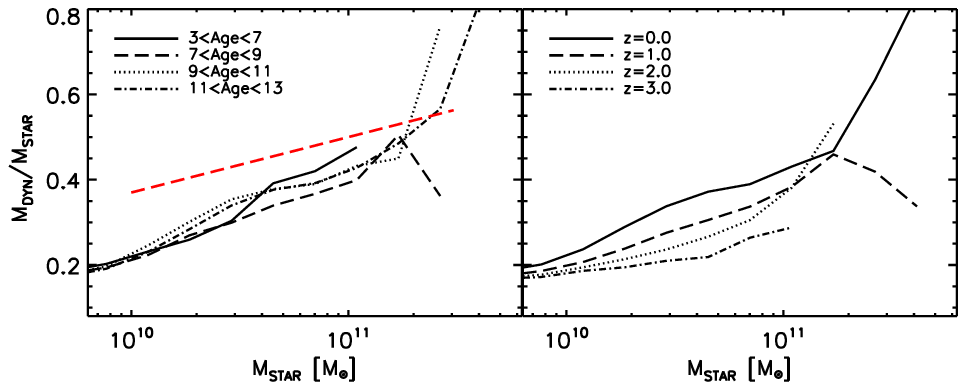}
    \caption{Predicted ratio of dynamical mass to stellar mass
    as a function of Age at $z=0$ (\emph{left}) and at different
    redshifts (\emph{right}). The \emph{red long-dashed} line is the tilt
    measured by \citet{ShankarBernardi}.
    The tilt of the fundamental plane is predicted to be rather independent
    of galaxy age at $z=0$, and to flatten out at $z>0$.}
    \label{fig|FPtilt}
\end{figure*}

\section{The tilt of the Fundamental Plane}
\label{subsubsec|FP}

Early-type galaxies obey a tight scaling relation
linking \sis, \mstar, and \re, the so-called fundamental plane (FP),
as expected from basic virial arguments \citep[e.g.,][]{DDFP}.
If the mass-to-light ratio is constant at all masses, i.e., $L\propto M_{\rm dyn}$,
from the virial relation one would then simply expect $L\propto \sigma^2 R_e$.
However, the observed FP relation has small but significant departures
from the theoretical predictions, suggesting that $M_{\rm dyn}/L$
has a non trivial dependence on mass. This ``tilt'' of the FP is
possibly a consequence of stellar effects and/or progressive
difference in the inner dark matter content of the most massive galaxies \citep[e.g.,][]{Labarbera08,Tortora09,LaBarbera10,Napolitano10,Graves10,Tortora12}.

\citet{ShankarBernardi} showed from a large sample of SDSS early-type galaxies
that the ratio $M_{\rm dyn}/\mstare \propto \mstare^\gamma$, with $\gamma \approx 0.13$,
i.e., it increases with increasing $\mstare$, in a way approximately independent of the age of the galaxies.
The left panel of Figure~\ref{fig|FPtilt} shows
the predicted ratio of dynamical mass to stellar mass as a function stellar
mass for galaxies of different (mass-weighted) age at $z=0$ and with $B/T>0.9$.
In agreement with the SDSS data, the model correctly predicts a positive tilt
in the FP relation very similar to the observed one (red long-dashed line), and
with weak dependence on age. The parametrization adopted
in Eq.~(\ref{eq|CovingtonSigma}) naturally induces a tilt because
lower mass galaxies have proportionally smaller sizes, therefore
lower dark matter fractions within \rh, thus proportionally lower velocity dispersions.
The right panel of Figure~\ref{fig|FPtilt} shows that the tilt is predicted
to decrease at higher redshifts because galaxies at higher redshifts
are associated to less massive dark matter hosts and also more compact
galaxies, thus, again, proportionally lower dark matter fractions within \rh.
The scatter of FP is measured to be small, thus being
an important quantity to compare models with. However, a detailed study
of the FP scatter is beyond the scope of this brief Appendix, and we will postpone it to future work.

\section{The connection with the central Black Hole: evolution in scaling relations}
\label{sec|mbhrel}

Dynamical observations have revealed that super-massive Black Holes (BHs)
are ubiquitous at the centres of most, if not all, local massive,
bulge-dominated galaxies, with their mass $\mbhe \sim 10^6-10^9 \, \msune$,
tightly correlated with the mass and velocity dispersion of the host bulge
\citep[see][for recent reviews]{FerrareseFord,ShankarReview}. It is
therefore natural that constraining the evolution of massive spheroidal
systems contemporarily implies understanding the origin and evolution of
BHs and of such tight scaling relations.

The Munich SAM self-consistently follows the evolution of BHs
during the hierarchical evolution of galaxies.
The model assumes that a fixed fraction of the cold gas is destabilized during merger
events and feeds the central BH (see, e.g., \citealt{Marulli08} for details).
A parameter controls how much gas mass is funnelled onto the central
BH and it is fine-tuned to reproduce the local \mbh-\mstar\ relation.
The model has been found to be consistent with AGN luminosity functions and
quasar clustering at different redshifts and luminosities
\citep[see details in][]{Marulli08,Marulli09,Bonoli09,Bonoli10}.

\begin{figure*}
    \includegraphics[width=15truecm]{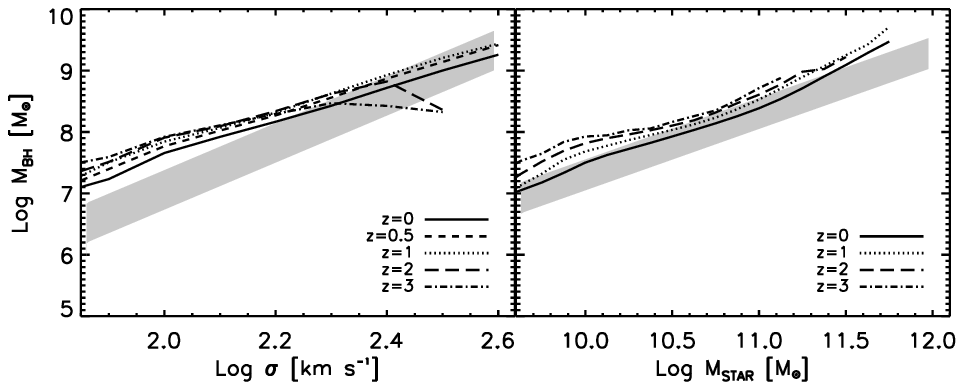}
    \caption{\emph{Left panel}: Predicted \mbh-\sis\ relation at different redshifts, as labelled,
    for a model with a dark matter mass-dependent \sis\ for
    galaxies with $B/T>0.9$. The \emph{grey stripe} indicates
    the fit by Tundo et al. (2007) with its intrinsic scatter.
    \emph{Right panel}: Predicted \mbh-\mstar\ relation at the same different
    redshifts for galaxies with $B/T>0.9$. The \emph{grey stripe} shows a
    linear relation of the type $\mbhe = 2\times 10^{-3}\, M_{\rm bulge}$ with some scatter,
    indicative of what suggested by a variety of local data.}
    \label{fig|MbhMstar}
\end{figure*}

The left panel of Figure~\ref{fig|MbhMstar} shows the \mbh-\sis\ relation
at different redshifts, as labelled,
for a model with \sis\ computed with Eq.~(\ref{eq|CovingtonSigma}).
For reference, the grey stripe indicates the fit by \citet{Tundo07}
with its intrinsic scatter. The model produces a reasonable match to the data
and we have checked that neglecting any halo mass dependence in
\sis\ would have produced a flattening similar to the one observed for
the \mstar-\sis\ relation in Figure~\ref{fig|MstarSigma}.

Interestingly, at variance with the evolution found for the \mstar-\sis\ relation,
the model predicts a positive evolution for the \mbh-\sis\ relation, i.e.,
comparable or higher BH masses at higher redshifts at fixed velocity dispersion.
This is in line with direct and indirect measurements of the \mbh-\sis\ at higher
redshift in quasar host galaxies
\citep[e.g.,][]{Shields06,Woo06,Woo08,Treu07,ShankarMsigma,GaskellMbhSigma,Bennert11}.
Though BH host galaxies get more compact and thus possess higher
velocity dispersions at higher redshifts, the redshift evolution
in the \mbh-\sis\ relation shows the opposite trend.

The reason behind the opposite time behaviour of the two relations
can be understood by looking at
the right panel of Figure~\ref{fig|MbhMstar}, which shows the
predicted \mbh-\mstar\ relation for the same early-type galaxy subsample.
The grey stripe is a linear relation of the type $\mbhe = 2\times 10^{-3}\, M_{\rm bulge}$
with some scatter, indicative of what suggested by a variety of local data
\citep[e.g.,][]{Marconi03,HaringRix}.
The overall agreement with the data at $z=0$ in the normalization
and slope of the \mbh-\mstar\ relation is
mostly a simple consequence of the underlying model that assumes both
BH accretion and star formation rate to be proportional to cold gas reservoir.
A genuine prediction of the model is
instead that the $\mbhe/\mstare$ ratio evolves at fixed
stellar mass, possibly in a mass dependent way, increasing
by a factor of a few at higher redshifts \citep[see also][]{Croton06b}, consistently with
what derived by many groups \citep[e.g.,][]{Decarli10}.
Thus the model predicts higher velocity dispersions but also
more massive BHs at fixed stellar mass at higher redshifts,
in a way to erase or even reverse the predicted evolution
in the \mbh\ - \sis\ relation, in good agreement with the data.
In other words, while most of the BH mass is accreted during
the high-$z$ gas-rich merger phase, the growth of the stellar mass
and velocity dispersion of the host spheroid is prolonged to later times.

\label{lastpage}
\end{document}